\pdfoutput=1

\documentclass[12pt,a4paper,notitlepage,oneside]{book}

\usepackage{csquotes}
\usepackage{graphicx}
\usepackage{amsmath}
\usepackage{apalike}

\newcommand{\bs}[1]{{\boldsymbol{{#1}}}}

\usepackage{abstract_def}

\usepackage{titling}
\usepackage{authblk}
\title{Analysing Multiple Epidemic Data Sources}
\author[1]{Daniela De Angelis}
\author[1]{Anne M. Presanis}
\affil[1]{MRC Biostatistics Unit, University of Cambridge}
\date{\vspace{-5ex}}

\begin{document}
\maketitle

\framebox{
  \begin{minipage}{0.9\linewidth}
    This manuscript is a preprint of a Chapter to appear in the
    \textit{Handbook of Infectious Disease Data Analysis}, Held, L.,
    Hens, N., O'Neill, P.D. and Wallinga, J. (Eds.). Chapman \& Hall/CRC,
    2018. Please use the book for possible citations.
  \end{minipage}
}\\[2ex]

\begin{chapabstract} 
  \small{
\noindent 
Evidence-based knowledge of infectious disease burden, including
prevalence, incidence, severity and transmission, in different
population strata and locations, and possibly in real time, is crucial
to the planning and evaluation of public health policies. Direct
observation of a disease process is rarely possible. However, latent
characteristics of an epidemic and its evolution can often be inferred
from the synthesis of indirect information from various routine data
sources, as well as expert opinion. The simultaneous synthesis of
multiple data sources, often conveniently carried out in a Bayesian
framework, poses a number of statistical and computational challenges:
the heterogeneity in type, relevance and granularity of the data,
together with selection and informative observation biases, lead to
complex probabilistic models that are difficult to build and fit, and
challenging to criticize. Using motivating case studies of influenza,
this chapter illustrates the cycle of model development and criticism in the context of Bayesian evidence synthesis, highlighting the challenges of complex model building, computationally efficient inference, and conflicting evidence.
}
  
\end{chapabstract}
\thispagestyle{empty}
\thispagestyle{empty}
\clearpage
\setcounter{page}{1}
\chapter{Analysing Multiple Epidemic Data Sources}

\section{Introduction}\label{intro}

\begin{displayquote}
``A catalogue of the number of deaths induced by the major
epidemics of historical time is staggering, and dwarfs the total
deaths on all battlefields.'' \cite{Anderson1991}
\end{displayquote}
This quote sets the problem of infectious diseases in perspective. Although major historical threats have been defeated \cite{Heesterbeek2015}, new emerging ones continue to challenge humans. It is not surprising that increasing effort has been made by policy makers to assess and anticipate the consequence of epidemics. Evidence-based knowledge of disease burden, including prevalence, incidence, severity, and transmission, in different population strata, in different locations and, if feasible, in real time, is becoming progressively key to the planning and evaluation of public health policies \cite{Heesterbeek2015}. Direct observation of a disease process is hardly ever possible. However, retrospective and prospective estimation of the key aspects of burden listed above is feasible through the use of indirect information collected in administrative registries. The previous chapters in this handbook (and the rich literature that exists, e.g. \cite{Heesterbeek2015,Birrell2017} and references therein) provide plenty of examples of how surveillance information, together with statistical models, can be used to reconstruct the disease process underlying the pattern of the observed data, infer the unobserved (latent) characteristics of the epidemic and forecast its evolution.

Here we focus, in particular, on statistical inference that makes simultaneous use of multiple data sources, including different streams of surveillance data, ad hoc studies and expert opinion.

This `evidence synthesis' approach is not new in medical statistics. Meta-analysis and network meta-analysis are well established approaches to combine data from studies of similar design, typically clinical trials \cite{Borenstein2009}. The idea has been generalised in the areas of medical decision-making \cite{Eddy1992}, technology assessment \cite{Spiegelhalter2004,Welton2012} and epidemiology (e.g. \cite{AdesSutton2006}) to assimilate data from sources of different types and/or studies of different designs and is becoming popular in other scientific fields, as modern technologies enable the collection and storage of ever increasing amounts of information (e.g. \cite{Wheldon2015,Clark2017}). For infectious disease, in the last ten years there has been a proliferation of papers employing multiple sources of information to reconstruct characteristics of epidemics of blood borne and respiratory diseases, including estimation of prevalence (e.g. HIV \cite{DeAngelis2014a}, HCV \cite{Harris2012,McDonald2014}, campylobacteriosis \cite{Albert2011}), severity (e.g. \cite{Presanis2014,Shubin2014}), incidence (e.g. toxoplasmosis \cite{WeltonAdes2005}, influenza \cite{McDonald2014a} and pertussis \cite{McDonald2015}) and transmission (e.g. influenza \cite{Birrell2011,Dorigatti2013}).

The use of multiple data sources poses a number of statistical and computational challenges: the combination of various sources, most likely affected by selection and informative observation biases and heterogeneous in type, relevance and granularity, leads to probabilistic models with complex structures, difficult to build and fit and challenging to criticize \cite{DeAngelis2014}.

In this chapter, we will use motivating case studies of influenza to introduce the evidence synthesis setting in infectious diseases, illustrate the building and fitting of relevant models, and highlight the opportunities offered and the challenges posed by the multiplicity of sources. The models we will concentrate on are typically Bayesian, as this framework offers a natural setup for the synthesis of information. 

The chapter is organised as follows: in Section \ref{sec:egs}, we describe our motivating examples; in Section \ref{sec:evsynIntro}, the generic framework for evidence synthesis for infectious diseases is introduced; the models developed for the chosen examples are presented in Sections \ref{sec:flusev} and \ref{sec:flutrans}; Section \ref{sec:challenges} is devoted to the challenges encountered in the building, fitting and criticism of models that incorporate multiple sources; we conclude with a final discussion in Section \ref{sec:discuss}.

\section{Motivating example: influenza\label{sec:egs}}
Public health responses aimed at mitigating the impact of an outbreak need reliable (and prompt) assessment of the likely severity and spread of the infection. This understanding is particularly key when a new pathogen emerges, potentially causing a pandemic, for example a new influenza strain as in 2009 \cite{Lipsitch2009} or more recently, the zika \cite{Kucharski2016} and ebola \cite{Camacho2015} outbreaks.
 
We will use examples of influenza severity and transmission estimation, in particular referring to the 2009 A/H1N1 pandemic in the United Kingdom, to illustrate how, in the absence of ideal information, estimation of severity and transmission can be carried out by using data from a multiplicity of sources.

\subsection{Severity \label{sec:sevIntro}}

\begin{figure}
\centering
\includegraphics[width = 0.8\textwidth]{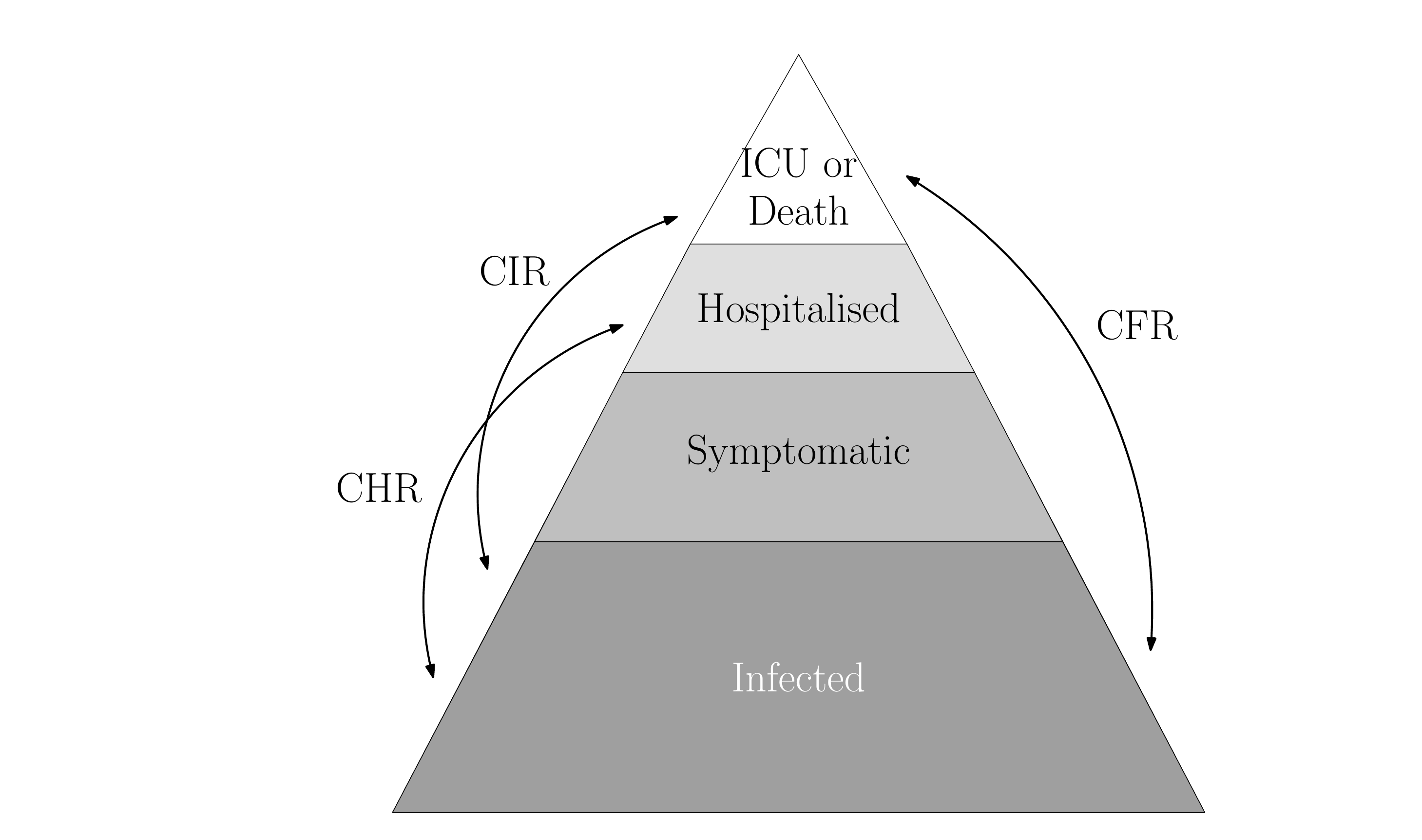}
\caption{Severity of influenza as a ``pyramid'': infected individuals progress from asymptomatic infection (`I') to symptomatic infection (`S'), hospitalisation (`H'), ICU-admission (`ICU') and/or death (`D'). Case-severity risks, i.e. the case-hospitalisation (CHR), case-ICU-admission (CIR) and case-fatality (CFR) risks, are defined as probabilities of a severe event given infection.}
\label{fig:pyramid}
\end{figure}
The severity of an infectious disease, such as influenza, can be thought as a ``pyramid'' (Figure \ref{fig:pyramid}), where with increasing severity, there are fewer and fewer infections. A proportion of infected individuals progress to symptoms, then to hospitalisation, and the most severe end-points of either intensive care (ICU) admission or death. Severity is usually expressed as ``case-severity risks'' (CSRs), {\it i.e.} probabilities that an infection leads to a severe event such as being hospitalised, admitted to ICU, or dying. Quantification of such risks is necessary both prospectively, in the midst of an outbreak, to understand the severity and likely burden on health-care of an ongoing epidemic; and retrospectively, to assess the severe burden of the particular strain responsible for the epidemic and the adequacy of any public health response during the outbreak, and to inform responses in future outbreaks. However, such CSRs are challenging to directly observe, requiring therefore estimation. 

Prospectively, estimation would require a cohort of cases, {\it i.e.} individuals with laboratory-confirmed influenza, to be followed up over time. However, a representative sample of those who are infected is almost impossible to recruit,  particularly as infections that are asymptomatic are less likely to be observed in health-care data than symptomatic infections. Even if it were possible, prospective estimation would have to account appropriately for censoring, as the end points of interest might take time to occur. 

For retrospective estimation, censoring may not be an issue, however, differential probabilities of observing cases at different levels of severity (ascertainment/detection probabilities) may lead to biases \cite{Lipsitch2009}. `Multiplier' methods \cite{Reed2009,Shrestha2011,Reed2015} therefore have been proposed when individual-level survival-type data are not available, combining aggregate case numbers at different levels of severity ({\it e.g.} surveillance data on sero-prevalence {\it i.e.} the proportion of blood samples testing positive for influenza; general practice (GP) consultations for influenza-like-illness (ILI); hospital/ICU admissions; mortality) to obtain estimates of the CSRs. These methods account for the ascertainment/detection biases suffered by aggregate surveillance data, through multiplication by inverse proportions detected, with informal uncertainty quantification using Monte Carlo forward simulation (e.g. \cite{Reed2009,Shrestha2011}). 

Hybrid methods, that combine hierarchical models with multiplier methods in different stages, have appeared recently \cite{Reed2015}. However, multiplier methods to estimate severity were first formalised using Bayesian evidence synthesis, to simultaneously account for uncertainty, prior information on some ascertainment/detection biases, and censoring \cite{Presanis2009,Presanis2011,Presanis2014}. The uncertainty inherent in each data source, together with prior uncertainty, is propagated formally through to posterior estimates of the CSRs. The estimated CSRs are derived as products of probabilities of being at one severity level conditional on being at lower severity level (Figure \ref{fig:pyramid}).

\begin{figure}
\includegraphics[width = \textwidth]{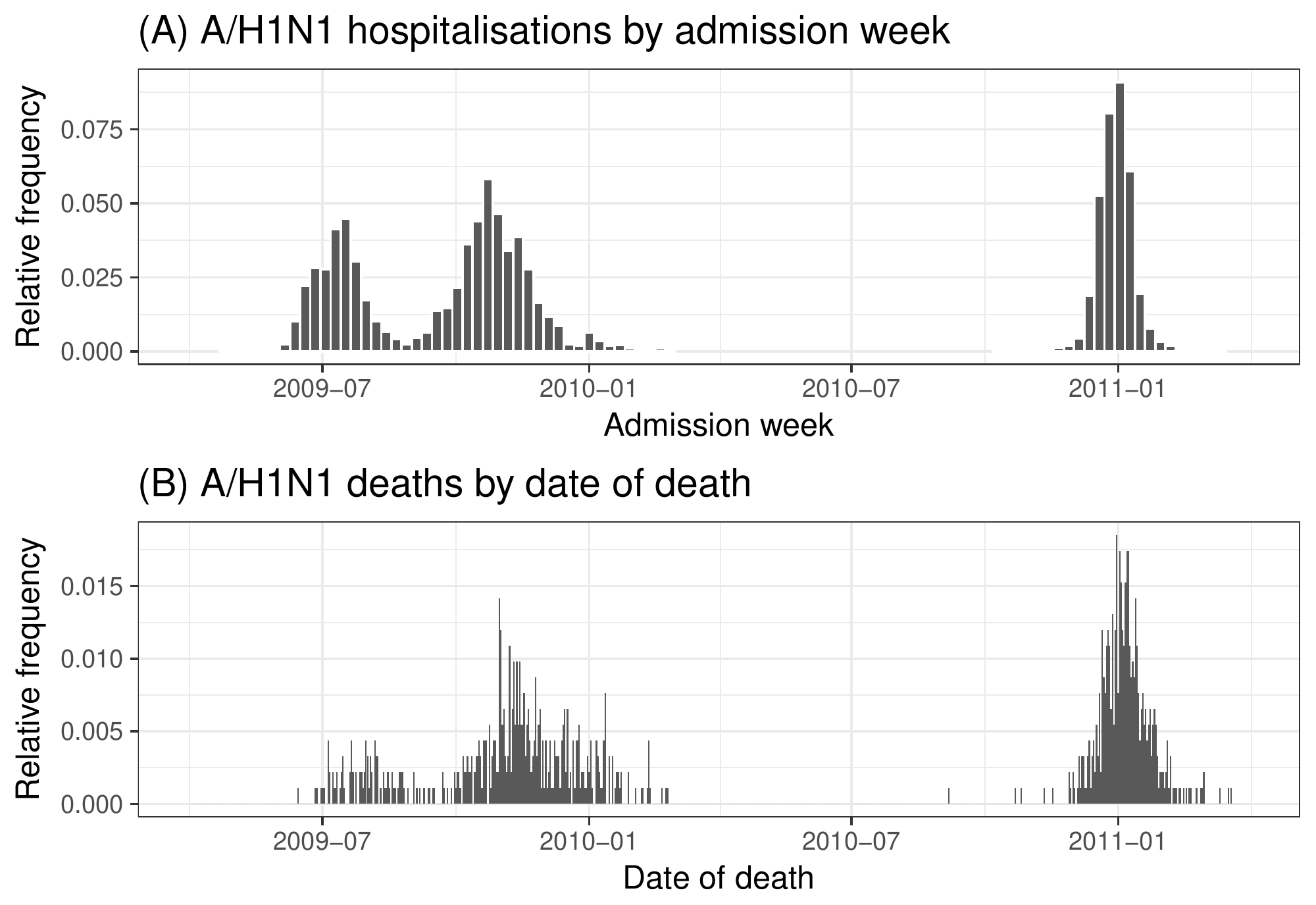}
\caption{Frequency of observed severe events relative to the total number of events over the period April 2009-March 2011 of the three waves of infection, among individuals with confirmed A/H1N1 pandemic influenza: (A) weekly hospitalisations; (B) daily deaths.}
\label{fig:obsSevereCounts}
\end{figure}

\begin{figure}
\centering
\includegraphics[width = \textwidth]{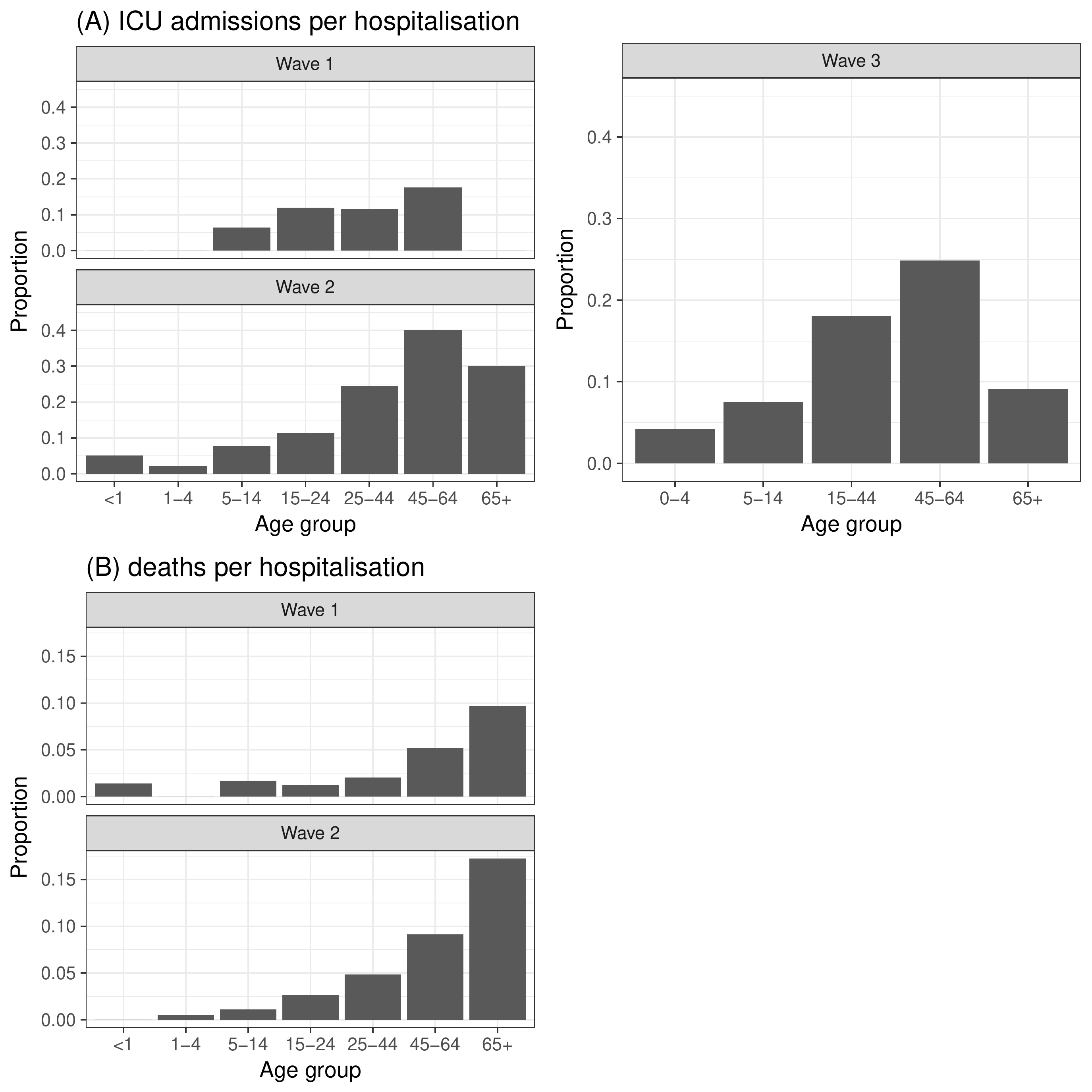}
\caption{Proportion of observed individuals hospitalised with confirmed A/H1N1 pandemic influenza who experienced severe events during the three waves of infection: (A) ICU admissions per confirmed A/H1N1 hospitalisation, by age and wave; (B) deaths per confirmed A/H1N1 hospitalisation, by age and wave.}
\label{fig:obsSeverePropn}
\end{figure}

Such an evidence synthesis is presented in \cite{Presanis2014}, for the A/H1N1 pandemic in England during each of three waves experienced in summer 2009, the 2009/2010 and 2010/2011 seasons. The available data sources include: (i) cross-sectional sero-prevalence data from laboratory-tested residual sera from patients after diagnostic testing for various (non-respiratory) conditions. These data, over time, inform changes in the proportion of the population exposed to influenza strains, i.e. the population level of immunity, and hence indirectly inform incidence; (ii) estimates of numbers symptomatic based on potentially under-ascertained GP consultations for ``influenza-like-illness'' (ILI) and corresponding data on the proportion of nasopharyngeal swabs from individuals with ILI that test virologically positive for the A/H1N1pdm strain; (iii) under-ascertained retrospective and prospective daily hospital admissions from 129 hospital trusts in the first two waves, and a sentinel set of 23 trusts in the third wave (Figure \ref{fig:obsSevereCounts}(A)); (iv) under-ascertained numbers of deaths occurring in individuals with confirmed A/H1N1 infection (Figure \ref{fig:obsSevereCounts}(B)). Each source poses a number of challenges, in addition to the above-mentioned ascertainment/detection biases. The sero-prevalence data, although available for all three waves, in the second and third waves, does not allow separation of individuals with antibodies in response to vaccination from infected individuals. Point estimates of the number symptomatic from the Health Protection Agency (HPA) are only available in the first two waves, with an informal ``uncertainty range'' from sensitivity analyses. For the third wave, such estimates are instead obtained from a joint regression model of the GP ILI and virological positivity data, based on a much smaller sentinel set of general practices than the more comprehensive sentinel system used in the first two waves. Both the GP and positivity datasets are required, to disentangle ILI consultations due to ``background'' consultations for other respiratory illness from actual influenza consultations. The switch from a comprehensive to a sentinel hospital system between the second and third waves results in sparser data and changes in the age groups recorded (Figure \ref{fig:obsSeverePropn}), particularly affecting the number of severe outcomes (ICU admissions and deaths) reported in the hospitalisation data: no deaths are observed in the third wave. This sparsity requires the use of an additional ICU data source, which poses its own challenge. The system measures prevalent cases present in ICU, rather than incident or cumulative incident ICU admissions, and hence requires a model of the process of admissions and discharges to obtain estimates of cumulative admissions. 

None of the data sources on their own can provide an estimate of all CSRs of interest. However, by combining them all in a Bayesian evidence synthesis, the challenges described above can be resolved to derive the necessary severity estimates, as presented in Section \ref{sec:flusev}. 

\subsection{Transmission} \label{sec:transIntro}

Understanding the dynamics of an infectious disease amounts to estimation of the rate at which it spreads and the factors that are contributing to its spread. To acquire such knowledge, mechanistic transmission models are used \cite{Anderson1991}, expressed by differential equations describing the disease dynamics resulting from the interaction between individuals at different disease stages. For instance, in the susceptible-infectious-recovered (SIR) model new infections are generated through the contact between susceptible and infectious individuals, {\it i.e.} in state S and I, respectively. For influenza, and other respiratory infections, the relevant contact is between different age groups, since school-age children and their interactions with both other children and adults are known to be key drivers of transmission \cite{Cauchemez2008}. Historically, studies of influenza transmission have been carried out either by simulation \cite{Heesterbeek2015} or by estimating the parameters of transmission models using direct information from a single time series of disease endpoints, such as confirmed cases (e.g. \cite{Cauchemez2008}).

However, in recent years the need and potential of combining data from multiple sources to infer latent characteristics of epidemics has been increasingly recognised. For influenza, in particular, since the 2009 A/H1N1 influenza pandemic, this recognition resulted in the development of a number of transmission models using data from either multiple surveillance time series ({\it e.g.} \cite{Birrell2011,Dorigatti2013,Baguelin2013,teBeest2014,Shubin2016,Birrell2016}) or a combination of surveillance and phylogenetic data ({\it e.g.} \cite{Ypma2012,Ratmann2012,Jombart2014}). The integration of different sources of evidence can ensure identification of interpretable parameters in transmission models and a more comprehensive description of the evolution of an outbreak \cite{DeAngelis2014}.

\begin{figure}
\includegraphics[width = \textwidth]{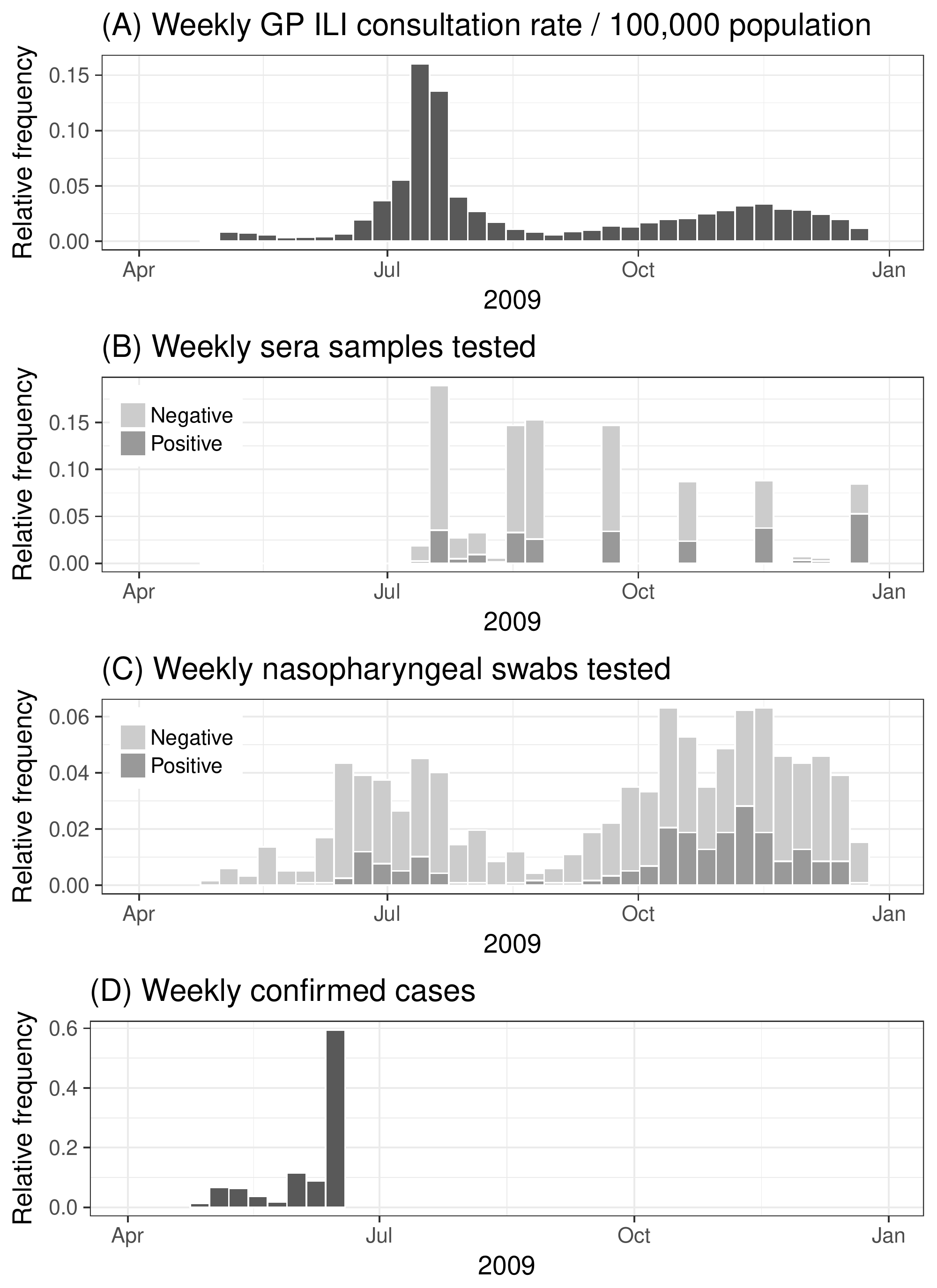}
\caption{Four data streams used in the transmission model of \cite{Birrell2011}: (A) GP ILI consultations; (B) serological positivity; (C) virological positivity; (D) confirmed cases. Each plot shows the frequency of events (consultations per 100,000 population, positive \& negative sera samples, positive \& negative swabs, confirmed cases respectively) relative to the total number of respective events over the period April 2009-December 2009 of the first two waves of infection.}
\label{fig:obsFluTrans}
\end{figure}

An example is given in \cite{Birrell2011} where, in the absence of a complete time series of confirmed influenza cases, various data sources are used to estimate retrospectively transmission during the first two waves of pandemic A/H1N1 influenza infection. Figure \ref{fig:obsFluTrans} shows the data for the London region: (A) GP consultations for ILI from May to December 2009; (B) a series of cross-sectional samples from sero-prevalence surveys (see Section \ref{sec:sevIntro}); (C) virological data on nasopharyngeal positivity for A/H1N1 (again as in Section \ref{sec:sevIntro}); and (D) a limited time series of confirmed cases in the first few weeks of the outbreak, up till June 2009, when contact tracing ceased. As in Section \ref{sec:sevIntro}, GP consultation data are contaminated by individuals experiencing non-A/H1N1-related ILI, whose health-care seeking behaviour is highly influenced by governmental advice and media reporting. To reconstruct the underlying pattern of A/H1N1 infections, GP data had to be combined  with information on A/H1N1 virological positivity, on population immunity from the serological surveys, knowledge on the natural history of A/H1N1, including the probability of developing symptoms, and data on the propensity of patients with symptomatic infections to consult a GP \cite{Brooks-Pollock2011}.

\section{Bayesian evidence synthesis \label{sec:evsynIntro}}

The notion of evidence synthesis is intrinsic to the Bayesian philosophy of assimilating information, Bayes' theorem being the basis for the combination of prior and new evidence. Generalising the concepts of meta-analysis and network meta-analysis, the evidence synthesis described and used here combines information from different study designs, through complex hierarchical models \cite{AdesSutton2006,DeAngelis2014,DeAngelis2014a}.

\paragraph*{A useful graphical representation \label{sec:evsyn}}

Bayesian hierarchical models have a long history of being expressed as directed acyclic graphs (DAGs), encoding the dependency structure between variables in the model \cite{Lauritzen1996}.
\begin{figure}[h]
\centering
\includegraphics[width = 0.5\textwidth]{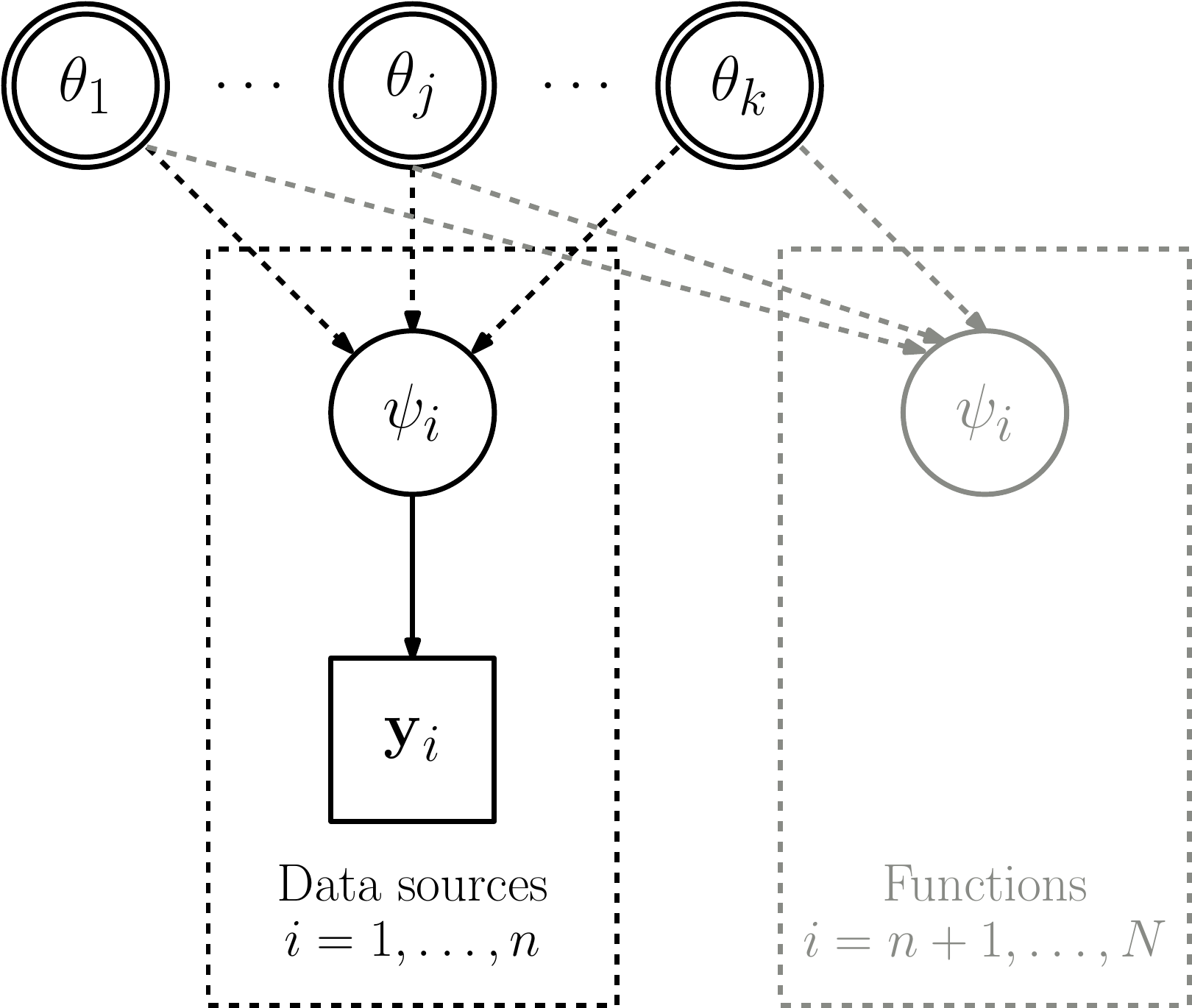}
\caption{Directed acyclic graph of a generic evidence synthesis model.}
\label{fig:dagES}
\end{figure}

A generic evidence synthesis model can be represented graphically as in Figure \ref{fig:dagES}. Square nodes represent observable quantities such as $\bs{y}_i$, whereas circles are latent quantities, such as $\psi_i$. Double circles such as $\theta_j$ are \emph{founder} nodes, i.e. parameters to which a prior distribution is assigned. Dashed rectangles, or ``plates'', represent repetition over indices, such as $i \in 1, \ldots, n$. Dependencies between variables are indicated by direct arrows, with solid and dashed arrows representing distributional (stochastic) and functional (deterministic) dependencies, respectively. The joint distribution of all quantities in the DAG is the product of the conditional distributions of each node given its direct parents. The aim of an evidence synthesis model such as Figure \ref{fig:dagES} is to estimate a set of $k$ \emph{basic} parameters $\bs{\theta} = \{\theta_1, \ldots, \theta_k\}$, based on a set of $n$ independent datasets $\bs{y} = \{\bs{y}_1, \ldots, \bs{y}_n\}$, where $n$ is not necessarily equal to $k$. Each dataset $\bs{y}_i, i \in 1, \ldots, n$ is assumed to inform a quantity (a \emph{functional} parameter) $\psi_i = \psi_i(\bs{\theta})$ that can be expressed as a deterministic function of the basic parameters. If $\psi_i \equiv \theta_j$ for some $j \in 1, \ldots, k$, $\bs{y}_i$ is said to \emph{directly} inform $\theta_j$. Otherwise, for $\psi_i = \psi_i(\bs{\theta})$, $\bs{y}_i$ \emph{indirectly} informs multiple parameters in the basic parameter set, in conjunction with all the other datasets. Further functional quantities $\psi_i = \psi_i(\bs{\theta}), i > n$ may be of interest to derive from the basic parameters, even if no data directly inform such functions. Assuming the independence of each dataset $\bs{y}_i, i \in 1, \ldots, n$ conditional on their common parents, the posterior distribution of the basic parameters $\bs{\theta}$ given the data $\bs{y}$ is
$$
p(\bs{\theta} \mid \bs{y}) \propto p(\bs{\theta}) \prod_{i=1}^n p(\bs{y}_i \mid \psi_i(\bs{\theta})) = p(\bs{\theta}) \prod_{i=1}^n p(\bs{y}_i \mid \bs{\theta}).
$$
Such an evidence synthesis model and DAG can clearly be extended both horizontally (more datasets) and vertically (hierarchical modelling).

\begin{figure}[h]
\centering
\includegraphics[width = \textwidth]{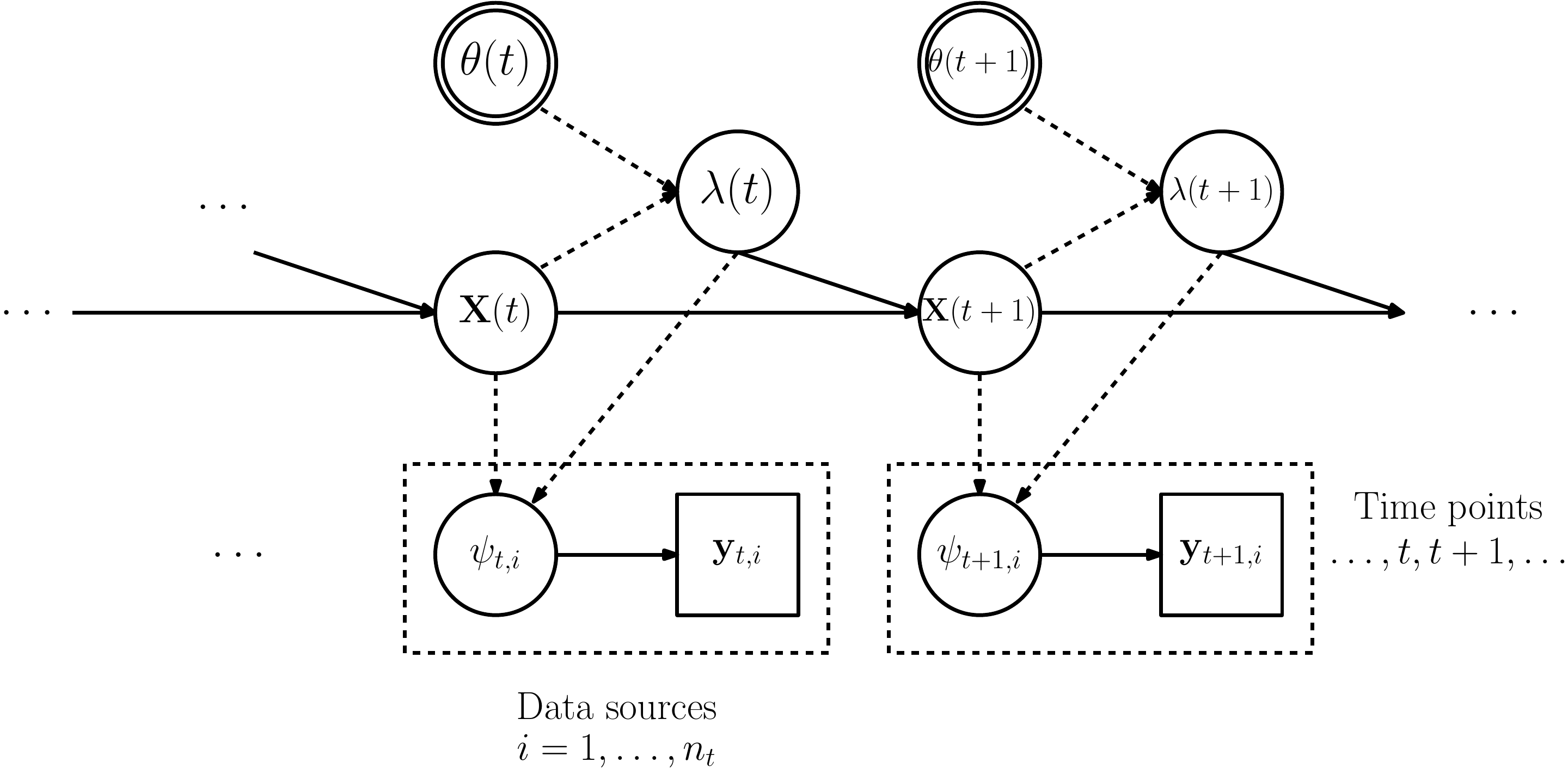}
\caption{Directed acyclic graph of a multi-state model representing a transmission model embedded in an
  evidence synthesis.}
\label{fig:dagMSM}
\end{figure}
Many evidence synthesis models used in the epidemic literature can be usefully represented graphically by DAGs \cite{AndersenKeiding2002,Birrell2016,Jackson2015}. Figure \ref{fig:dagMSM} is a generic DAG representation of a population-level multi-state disease transmission model where the vector-valued $\boldsymbol{X}(t)=(X_1(t),\ldots, X_K(t))$ corresponds to either the number or proportion of the population in each disease state ${1,\ldots, K}$ at time $t$. Movement between states is governed by transition rates $\boldsymbol{\lambda}(t)$, which are parameterised in terms of a collection of unknown basic parameters $\bs{\theta}(t)$, and the current state of the system $\bs{X}(t)$. If transmission is not explicitly modelled, the dependence of $\bs{\lambda}(t)$ on the states $\bs{X}(t)$, represented by the dashed arrow from $\bs{X}(t)$ to $\bs{\lambda}(t)$, is removed, and the model simplifies to a standard linear multi-state model. Typically, the $\boldsymbol{X}(t)$ and $\boldsymbol{\lambda}(t)$ are not directly observed. Instead, as in the simpler case of Figure \ref{fig:dagES}, observations ${\boldsymbol{y}_{t,i}, i=1, \ldots ,n_t}$ are available at time $t$, with each ${\boldsymbol{y}_{t,i}}$ informing a functional parameter $\boldsymbol{\psi}_{t,i} =\boldsymbol{\psi}_{t,i}(\boldsymbol{X}(t), \boldsymbol{\lambda}(t)$. These relationships may be stochastic dependencies, or more usually, deterministic functions. Again assuming that the ${\boldsymbol{y}_{t,i}}$ for $i=1, \ldots,n_t$ are independent conditional on their common parents, the likelihood of the data $\boldsymbol{y_t} = (\boldsymbol{y}_{t,i}, i = 1, \ldots, n_t)$ at time $t$ is expressed as the product 
$$
L(\boldsymbol{y_t} \mid \boldsymbol{X}(t),\boldsymbol{\lambda}(t)) = \prod_{i=1}^{n_t}
L(\boldsymbol{y}_{t,i} \mid \boldsymbol{\psi}_{t,i}(\boldsymbol{X}(t),
\boldsymbol{\lambda}(t))) .
$$
When the data are also conditionally independent over time, the likelihood of all the data $\boldsymbol{y} = (\boldsymbol{y}_{0,1}, \ldots,
\boldsymbol{y}_{t,n_t})$ given the basic parameters $\bs{\theta} = (\bs{\theta}_0, \ldots, \bs{\theta}_t, \ldots)$ is
$$
L(\boldsymbol{y} \mid \boldsymbol{\theta}) = \prod_t \prod_{i=1}^{n_t}
L(\boldsymbol{y}_{t,i} \mid \boldsymbol{\psi}_{t,i}(\boldsymbol{X}(t),
\boldsymbol{\lambda}(t))) = \prod_t \prod_{i=1}^{n_t}
L(\boldsymbol{y}_{t,i} \mid \bs{\theta}_t).
$$
The posterior distribution of $\bs{\theta}$ given the data is then $p(\bs{\theta} \mid \bs{y}) \propto L(\bs{y} \mid \bs{\theta}) p(\bs{\theta})$. Note that given the posterior distribution of the basic parameters and/or the states and transition rates, any function $\psi(\bs{\theta},\bs{\lambda},\bs{X})$, even if not directly observed, can be derived.

\section{Cross-sectional estimation: severity \label{sec:flusev}}

Static, cross-sectional models to estimate prevalence or cumulative incidence of (perhaps severe) disease are particular cases of the model in Figure \ref{fig:dagMSM}. The influenza severity estimation of \cite{Presanis2014} can be seen as three cross-sectional analyses, one for each wave of infection, where severity is expressed in terms of ratios of cumulative incidence of infection at different levels of severity (Figure \ref{fig:pyramid}). The model is also stratified by age group $a \in \{<1, 1-4, 5-14, 15-24, 25-44, 45-64, 65+\}$. The three timepoints $t \in \{0,1,2\}$ are not completely independent, as they share some parameters. 

\subsection{Model specification}

\subsubsection{First and second waves}
\begin{figure}[h]
\centering
\includegraphics[width = \textwidth, trim = 0 550 450 0, clip]{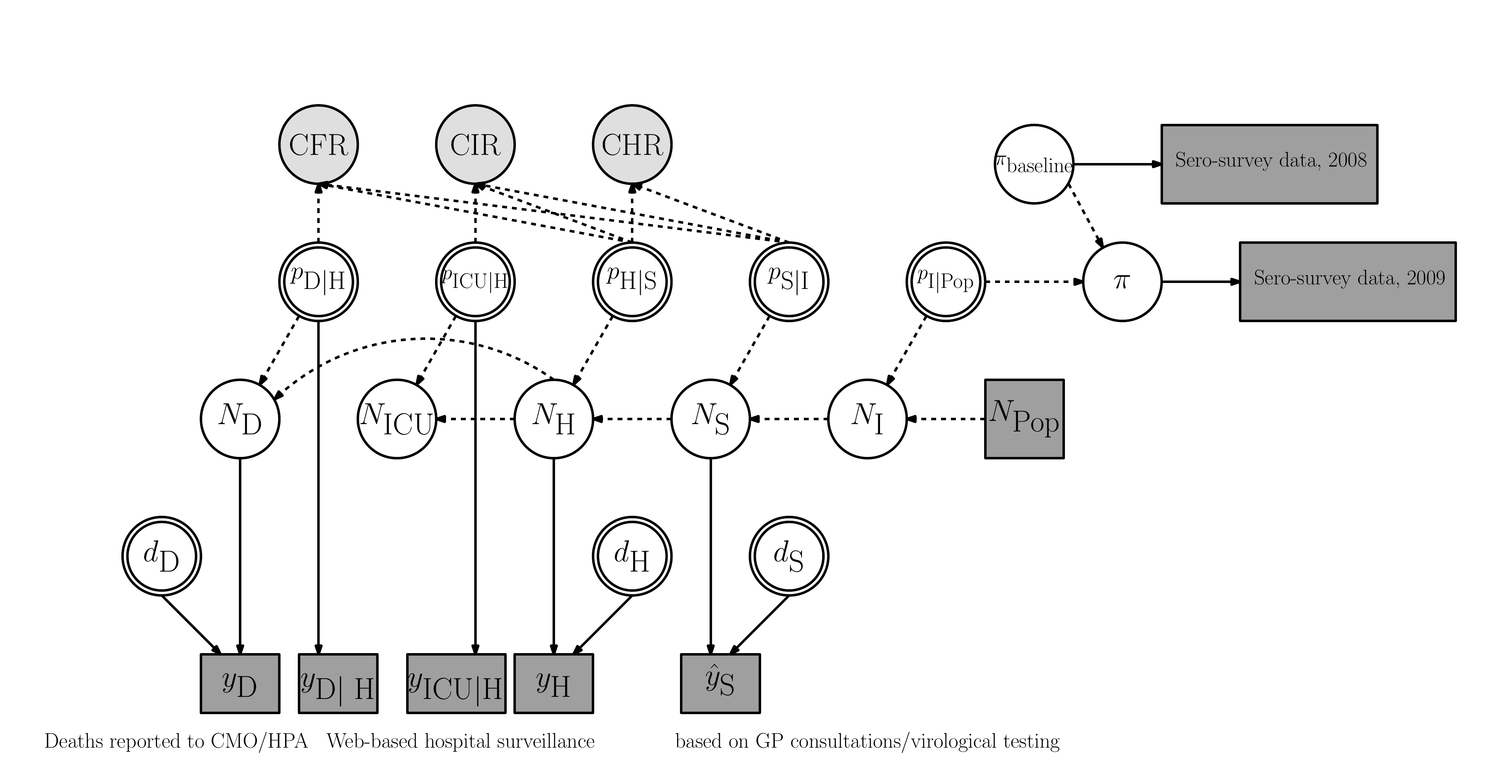}
\caption{DAG adapted from \cite{Presanis2014} representing the first wave of A/H1N1 pandemic influenza infection.}
\label{fig:dagWaves12}
\end{figure}

Figure \ref{fig:dagWaves12} displays the DAG corresponding to the first wave of infection, in summer 2009 ($t = 0$ in the notation of Figure \ref{fig:dagMSM}). The disease states $\boldsymbol{X}(0)$ correspond to the numbers $N_l$ of infections at each severity level $l = \{\textrm{I,S,H,ICU,D}\}$ (Figure \ref{fig:pyramid}). Note that age and wave/time indices have been omitted in the DAG and what follows, for brevity. 

\paragraph*{\bf Functional parameters}
Ideally, we would assume a nested binomial structure for the states $N_l \sim \textrm{Binomial}(N_m, p_{l \mid m})$ for a severity level $m$ lower than $l$ and for conditional probability $p_{l \mid m}$ of being a case at level $l$ given infection at level $m$. However, for computational reasons, we instead assume a mean parameterisation such that generically, the states $N_l$ are deterministic functions
$$
N_l = \left\lfloor p_{l \mid m} N_{m} \right\rfloor
$$
of the number $N_{m}$ of infections at a lower level $m$ of severity and the conditional probability $p_{l \mid m}$ of being a case at level $l$ given infection at level $m$. The total population $N_{\textrm{Pop}}$ is considered fixed. The case-severity risks of interest, \emph{i.e.} the case-hospitalisation (CHR), case-ICU admission (CIR) and case-fatality (CFR) risks (Figure \ref{fig:pyramid}), are products of component conditional probabilities:
\begin{align*}
\textrm{CHR} & = p_{\textrm{H} \mid \textrm{I}} = p_{\textrm{H}\mid \textrm{S}} \times p_{\textrm{S}\mid \textrm{I}} \\
\textrm{CIR} & = p_{\textrm{ICU}\mid \textrm{I}} = p_{\textrm{ICU}\mid \textrm{H}} \times \textrm{CHR} \\
\textrm{CFR} & = p_{\textrm{D}\mid \textrm{I}} = p_{\textrm{D}\mid \textrm{H}} \times \textrm{CHR}.
\end{align*}
The functional parameters $\boldsymbol{\psi}$ of Figure \ref{fig:dagMSM} are defined as the set $\boldsymbol{\psi} = \{N_l, l \in \{\textrm{I,S,H,ICU,D}\}\} \cup \{\textrm{CHR, CIR, CFR}\}$ of states and the corresponding case-severity risks.

\paragraph*{\bf Observational model} The observations, $y_i, i = 1, \ldots, n$, where $n = 7$ per age group, are either reported numbers of infections at different levels of severity ($\hat{y}_{\textrm S}, y_{\textrm H}, y_{\textrm D}$ in Figure \ref{fig:dagWaves12}), reported numbers of hospitalisations that lead to severe events ($y_{\textrm{ICU}\mid \textrm{H}}, y_{\textrm{D}\mid \textrm{H}}$), or the cross-sectional sero-prevalence data before and after the first wave. Each observation is a realisation of a binomial distribution, with probability parameters representing either: a detection probability $d_l$ with size parameter the counts of the numbers of infections $N_l$ at levels $l \in \{\textrm{S,H,D}\}$; a conditional probability $p_{\textrm{ICU} \mid \textrm{H}}$ or $p_{\textrm{D} \mid \textrm{H}}$ with size parameter the subset of observed hospitalisations $y_{\textrm{H}}$ where the final outcome $y_{\textrm{ICU}\mid \textrm{H}}$ or $y_{\textrm{D} \mid \textrm{H}}$ or discharge was observed; or a sero-prevalence, either before the first wave ($\pi_{\textrm{baseline}}$) or after the first wave ($\pi$), with size parameter the number of sera samples tested. The latter sero-prevalence data inform, indirectly, the infection attack rate $p_{\textrm{I} \mid \textrm{Pop}}$, via their difference. 

\paragraph*{\bf Basic parameters} The basic parameters $\boldsymbol{\theta}$ of Figure \ref{fig:dagMSM} are the set $\boldsymbol{\theta} = \{p_{l \mid m}, l,m \in \{\textrm{I,S,H,ICU,D}\}\} \cup \{d_l, l \in \{\textrm{S,H,D}\}\}$ of conditional and detection probabilities in Figure \ref{fig:dagWaves12}. Each probability is given an independent flat prior, apart from the symptomatic case-hospitalisation risk $p_{\textrm{H}\mid\textrm{S}}$ which is assigned an informative prior based on external data.

The second wave model is similar to that of the first wave, excluding only the sero-prevalence data, due to the challenges of disentangling vaccinated cases from true infections in the sero-samples in the absence of good data on vaccination in the dataset. 

\subsubsection{Third wave \label{sec:sevWave3}}
\begin{figure}[h]
\centering
\includegraphics[width = \textwidth, trim = 0 550 750 0, clip]{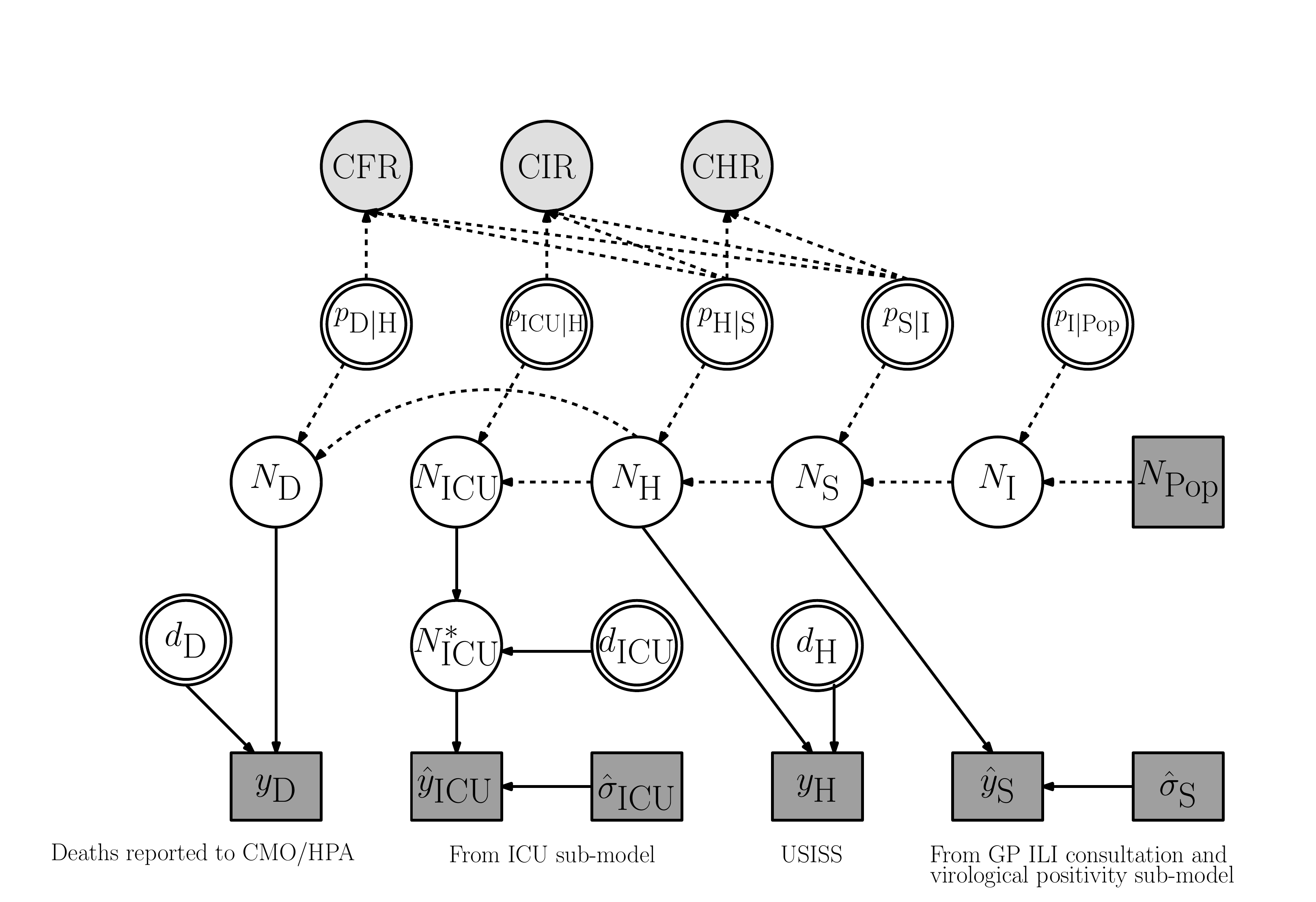}
\caption{DAG adapted from \cite{Presanis2014} representing the third
  wave of A/H1N1 pandemic influenza infection.}
\label{fig:dagWave3}
\end{figure}

\begin{figure}[h]
\centering
\includegraphics[width = 0.5\textwidth]{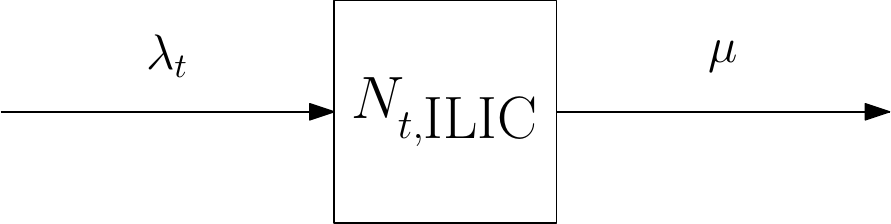}
\caption{Immigration-death process model for ILI cases in ICU. $\lambda_t$ is the daily rate of admissions to ICU, $N_{t,\textrm{ILIC}}$ the number of ILI cases present in ICU on day $t$, and $\mu$ the rate of exit (discharges or deaths), so that the expected length of stay in ICU is $1 / \mu$.}
\label{fig:ICUprocess}
\end{figure}

The third wave model differs more substantially (Figure \ref{fig:dagWave3}). Estimates of the number of symptomatic infections were derived from a joint Bayesian model of the GP consultation and virological positivity data from the smaller sentinel system, regressed on age group and time. This sub-model was fitted at a first stage, accounting for the probability of consulting a GP given symptoms, and giving posterior mean (sd) estimates $\hat{y}_{\textrm{S}}$ ($\hat{\sigma}_{\textrm{S}}$) on a log scale. In contrast to the first two waves, these estimates are not considered under-ascertained. They are therefore incorporated in the third wave model, at a second stage, via a likelihood term, assuming $\hat{y}_{\textrm{S}} \sim \textrm{Normal}(\log(N_{\textrm{S}}), \hat{\sigma}_{\textrm{S}})$.
Finally, since the hospitalisation data in the third wave (Figures \ref{fig:obsSevereCounts} and \ref{fig:obsSeverePropn}) are sparse, they lead to uncertain and under-ascertained estimates of the numbers hospitalised and the proportion of hospitalisations leading to ICU admission. Extra prevalence-type data on the number $N_{t,\textrm{ILIC}}$ of suspected ILI cases present in all ICUs in England are therefore incorporated, through a sub-model for these data that assumes entrances at rate $\lambda_t$ and exits at rate $\mu$ to/from ICU form an immigration-death stochastic process \cite{Presanis2014} (Figure \ref{fig:ICUprocess}). Note that this ICU sub-model is another example of a multi-state model as in Figure \ref{fig:dagMSM}, where the state is $N_{t,\textrm{ILIC}}$.
Since the observations are of influenza-like-illness rather than confirmed influenza, estimates of the ILI admission rate $\lambda_t$ are combined with virological positivity data from secondary care to obtain estimates of the cumulative number $y_{\textrm{ICU}}$ of new ICU admissions for A/H1N1 influenza. The posterior mean (sd) estimates $\hat{y}_{\textrm{ICU}}$ ($\hat{\sigma}_{\textrm{ICU}}$) from this sub-model, on a log scale, are then incorporated in the third wave model via a contribution to the likelihood:
\begin{equation}
\hat{y}_{\textrm{ICU}} \sim \textrm{Normal}(\log(N_{\textrm{ICU}}^*), \hat{\sigma}_{\textrm{ICU}}) \label{eq:ICUlogNorm}
\end{equation}
where $N_{\textrm{ICU}}^*$ is considered a lower bound for the total number of A/H1N1 ICU admissions, $N_{\textrm{ICU}}$, since the prevalent ICU case data cover only a portion of the time period of the third wave. This lower bound is implemented through a binomial assumption
$$
N_{\textrm{ICU}}^* \sim \textrm{Binomial}(N_{\textrm{ICU}}, d_{\textrm{ICU}})
$$
with probability parameter representing a detection probability $d_{\textrm{ICU}}$.

In contrast to the first two waves, priors for the conditional probabilities $p_{t=3,l\mid m}$ were expressed hierarchically by centering these probabilities, on a logit scale, on their respective second wave versions, $p_{t=2,l\mid m}$.

\subsection{Results}

\begin{figure}[h]
\centering
\includegraphics[width = \textwidth]{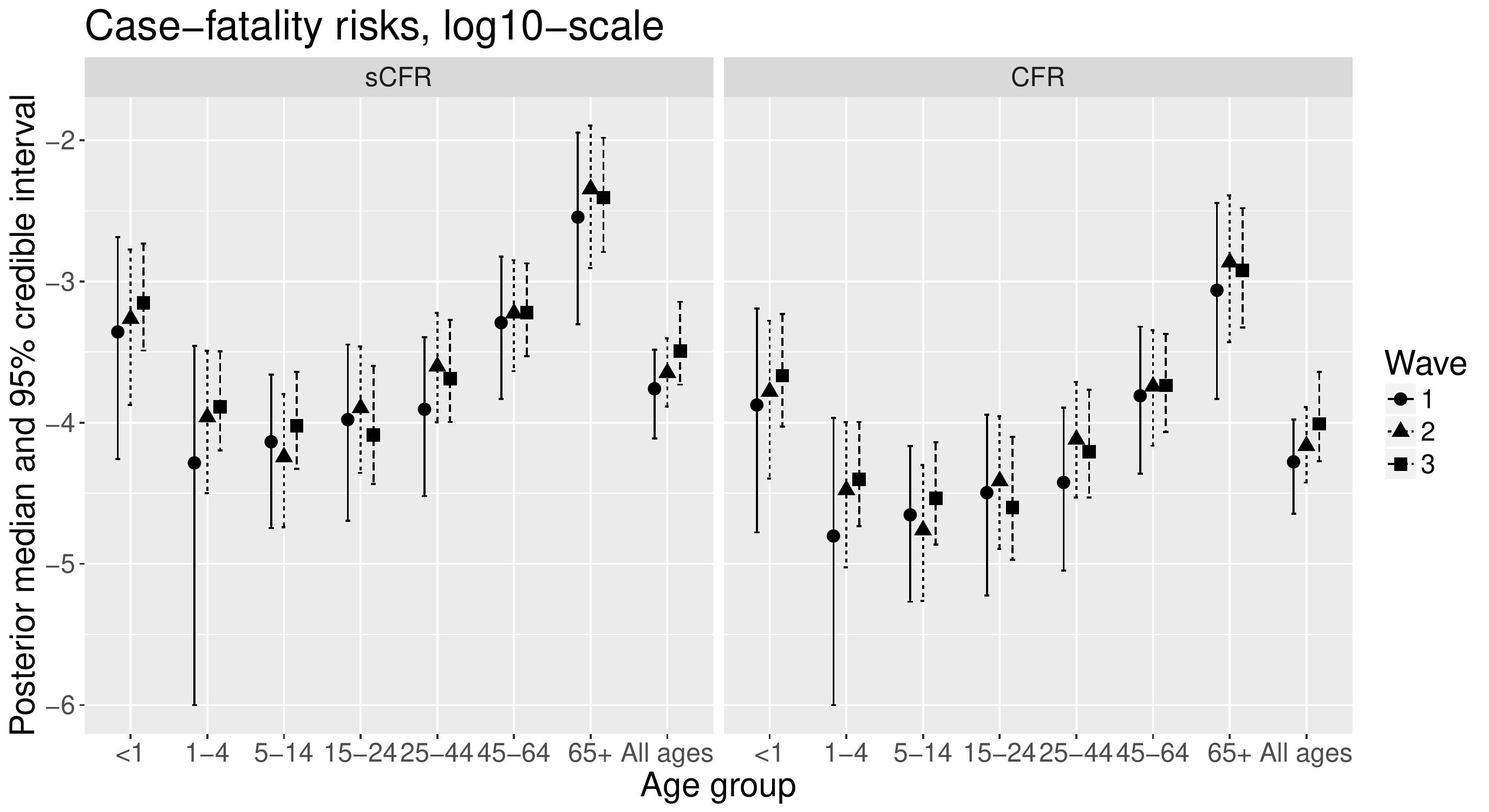}
\caption{Posterior median and 95\% credible intervals of symptomatic CFR and CFR, by wave and age group, log-scale.}
\label{fig:CFRsByWaveAge}
\end{figure}

\begin{figure}[h]
\centering
\includegraphics[width = \textwidth]{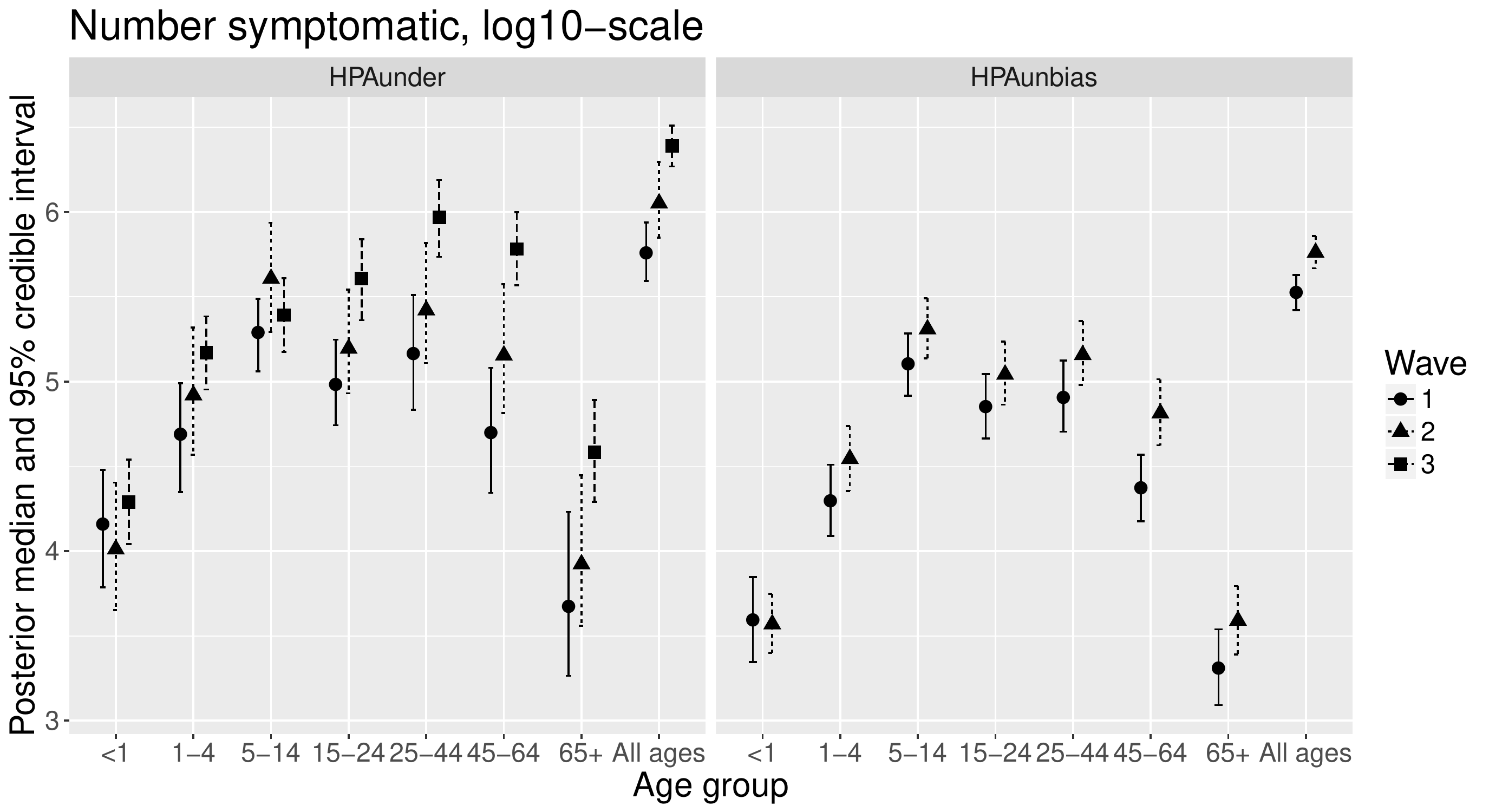}
\caption{Left: Posterior median and 95\% credible intervals of number of symptomatic infections, by wave and age group, log-scale. Right: Same posterior summaries, but for a sensitivity analysis assuming $d_{\textrm{S}} = 1$.}
\label{fig:NsymByWaveAge}
\end{figure}

Figure \ref{fig:CFRsByWaveAge} and the left-hand side of Figure \ref{fig:NsymByWaveAge} give posterior summaries of the symptomatic CFR (sCFR = $p_{\textrm{D}\mid \textrm{S}}$), CFR and number of symptomatic infections $N_{\textrm{S}}$, by age and wave. Note that for the first two waves of infection, the model giving these results assumes that the HPA-provided estimates of $N_{\textrm{S}}$ are under-estimates, with detection probability $d_{\textrm{S}}$. In \cite{Presanis2014}, discussion of the assumptions about the potential under-estimation in HPA estimates of the number symptomatic led to a number of sensitivity analyses. An initial unpublished sensitivity analysis assumed the HPA estimates were unbiased, \emph{i.e.} $d_{\textrm{S}} = 1$, giving the posterior estimates of $N_{\textrm{S}}$ on the right-hand side of Figure \ref{fig:NsymByWaveAge} for waves 1 and 2 only.

Notable patterns include: the `u'-shaped age distribution of the case-fatality risks and corresponding `n'-shaped age distribution of symptomatic cases; and an age shift towards older ages over the three waves of infection. The sensitivity analysis assuming $d_{\textrm{S}} = 1$ gives lower estimates of $N_{\textrm{S}}$ (and hence higher CSRs, not shown), but with a similar age pattern.

Further sensitivity analyses, as reported in \cite{Presanis2014}, suggested the key age patterns were more robust to prior choices than to the choice of bias/detection model for the HPA estimates.

\section{Temporal estimation: transmission \label{sec:flutrans}}

Compartmental mechanistic models, typically used to describe the process of disease transmission (see Section \ref{sec:egs} and Chapter ?), can also be represented as in Figure \ref{fig:dagMSM}. The disease states $\bs{X}(t)$ are of the classic Susceptible (S), Infected (I), Recovered (R) type and the transition rates $\bs{\lambda}(t)$ are functional parameters defined in terms of the current disease states. In particular, for the subset of  $\bs{\lambda}(t)$ representing incidence rates, we define $\bs{\lambda}_{\textrm{Inc}}(t) = f(\bs{X}_{\textrm{I}}(t))$ where $\bs{X}_{\textrm{I}}(t)$ is the current size of the infected and/or infectious states.

In \cite{Birrell2011}, the transmission of a novel pandemic A/H1N1 influenza strain among a fixed population stratified into $A$ age groups is estimated through the combination of a deterministic age-structured transmission model with disease and reporting models, describing disease transmission, progression and health-care seeking behaviour of infected individuals, respectively.

\subsection{Model specification}

\begin{figure}[!ht]
\centering
\includegraphics[width = \textwidth, trim = 200 0 300 0, clip]{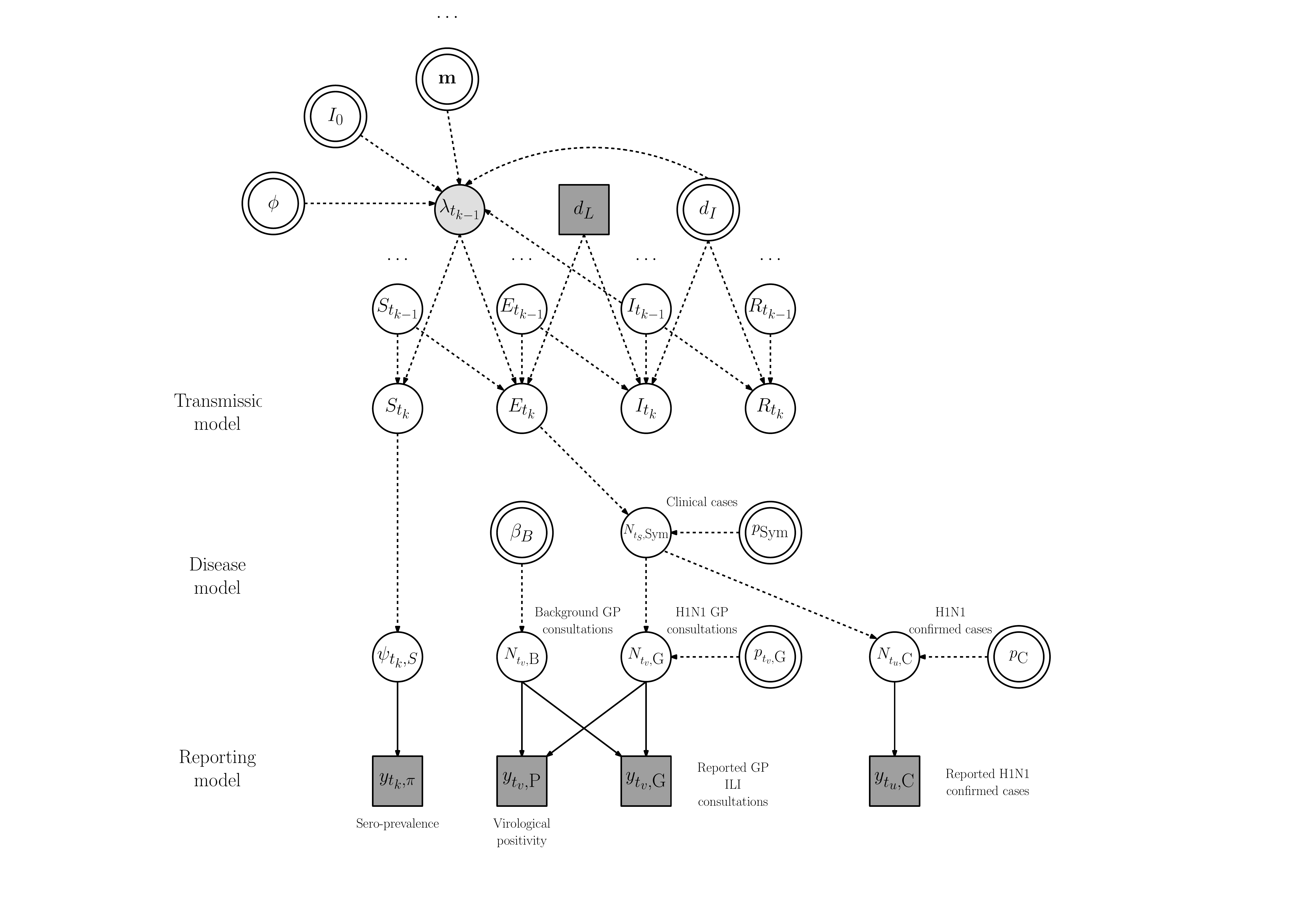}
\caption{DAG representing the transmission model of \cite{Birrell2011}.}\label{fig:fluTransDAG}
\end{figure}

\paragraph*{\bf Transmission model}
The transmission dynamics are governed by a system of differential equations of the type: 
\begin{equation}\label{eqn:determ.dynam}
\begin{split}
\frac{dS(t,a)}{dt} &= -\lambda(t,a)S(t,a)\\
\frac{dE(t,a)}{dt} &= \lambda(t,a)S(t,a) - \frac{1}{d_L}E(t,a)\\
\frac{dI(t,a)}{dt} &= \frac{1}{d_L}E(t, a) - \frac{1}{d_I}I(t, a)\\
\end{split}
\end{equation}
where $S(t,a)$, $E(t,a)$, $I(t,a)$ represent the number (or proportion) of the population of age group $a, (a=1,\ldots, A)$ in the $S$ (Susceptible), $E$ (Exposed) and $I$ (Infectious) states at time $t$ and $d_L$ and $d_I$ are the mean latent and infectious periods. Transmission is driven by the time- and age-varying rate $\lambda(t, a)$ at which susceptible individuals become infected. The system in \eqref{eqn:determ.dynam} is evaluated using an Euler approximation at times $t_k = k\delta t, k = 0, \ldots, K$, where the choice of $\delta t = 0.5$ days is sufficiently small that the probability of more than one change of state per period is negligible. Under this discretisation, at time $t_k$ the vector $(S_{t_k, a}, {E}_{t_k, a}, {I}_{t_k, a})$ gives the number of individuals in each state with the number of new infections in $[t_{k-1},t_k)$ being $\Delta_{t_{k},a} = \lambda_{t_{k - 1}, a}S_{t_{k - 1}, a}$, where
\begin{equation}
\lambda_{t_k, a} = 1 - \prod_{b = 1}^A \left\{ \left(1 - M_{t_k}^{(a, b)}R_{0}(\phi)/ d_I\right)^{{I}_{t_k, b}} \right\} \delta t. \label{eqn:foi}
\end{equation}
Here $R_{0}$ is the basic reproduction number, the expected number of secondary infections caused by a single primary infection in a fully susceptible population, often parameterised in terms of the epidemic growth rate $\phi$ \cite{Wearing2005}; and the time-varying mixing matrices $\boldsymbol{M}_{t_k}$,  express the pattern of transmission between age groups, with the generic entry $M_{t_k}^{(a, b)}$ being the relative rate of effective contacts between individuals of each pair of age groups $(a, b)$ at time $t_k$. The quantity $1 - M_{t_k}^{(a, b)}R_{0}(\phi) / d_I$ gives the probability of an individual in age group $a$ not being infected by an infectious individual in age group $b$ in the interval ${k+1}$. When raised to the power of all the infectious individuals in group $b$, the probability of not being infected by any individual in group $b$ is obtained. Taking the product over all age groups gives the probability of not being infected at all. This expression for $\lambda_{t_k,a}$ is known as the Reed-Frost formulation \cite{Ball1983}. The initial conditions of the system are determined by: parameter $I_0$, the total number of infectious individuals across all age groups at time $t_0$; an assumed equilibrium distribution of infections over the age groups; and an assumption of initial exponential growth that determines the relationship between the numbers in the four disease states. The mean latent period $d_L$ is taken as known, whereas the mean infectious period $d_I$ is a parameter to be estimated. Therefore, the dynamics of the transmission model \eqref{eqn:determ.dynam} depend on the \emph{basic} parameter vector $\bs{\theta}_T = (\phi, I_0, d_I, \boldsymbol{m})$, where $\boldsymbol{m}$ parameterise the mixing matrices $\boldsymbol{M}_{t_k}$. The transmission model is represented schematically in DAG format in Figure \ref{fig:fluTransDAG}. The dependency on age has been omitted in the DAG and in what follows, for brevity.

\paragraph*{\bf Disease progression and health-care seeking}
The newly infected individuals  $\Delta_{t_k}$, following an incubation time, develop ILI symptoms with probability $p_{\textrm{Sym}}$ (Figure \ref{fig:fluTransDAG}). With probability ${p}_{\textrm{C}}$, the symptomatic cases are virologically confirmed through contact tracing or hospitalisation in the early phase of the epidemic; and with time-varying probability ${p}_{t,\textrm{G}}$, symptomatic patients choose to contact a primary care practitioner (GP). These processes result in the (latent) number of symptomatic cases $N_{t_s,\textrm{Sym}}$, confirmed cases $N_{t_u,\textrm{C}}$ and GP consultations $N_{t_v,\textrm{G}}$, which can each be expressed as a convolution of the new infections $\Delta_{t_k}$ with the distribution of the time delay between infection and the relevant health-care event. For instance, the number $N_{t_v,\textrm{GP}}$ of GP consultations in the interval $[t_{v-1},t_v)$ is
\begin{equation}\label{eqn:convolution}
 N_{t_v,\textrm{GP}}= p_{Sym} p_{t_v,\textrm{G}} \sum_{k = 0}^v \Delta_{t_{k}} f(v-k)
\end{equation}
where the (discretised) delay  probability mass function $f(\cdot)$ accounts for both the time from infection to symptoms and the time from symptoms to GP consultation. The disease progression component of the model is specified by \emph{basic} parameters $\bs{\theta}_D = (p_{\textrm{Sym}},p_{\textrm{C}},p_{t_v,\textrm{G}})$ (Figure \ref{fig:fluTransDAG}) and, from equation (\ref{eqn:convolution}), the quantities $N_{t_u,\textrm{C}}$ and $N_{t_v,\textrm{G}}$ are complex functions of both $\bs{\theta}_T$ and $\bs{\theta}_D$.

\paragraph*{\bf Observational model} The goal here is to estimate the rate of infections $\lambda_{t_{k-1}}$ over time and predict the resulting burden on health-care facilities, through the estimation of the basic parameters $\bs{\theta} = \bs{\theta}_T \cup \bs{\theta}_D$ from observed data. As anticipated in Section \ref{sec:transIntro}, direct data on the number of new infections are not available. Therefore, a combination of a number of indirect evidence sources informing different aspects of the infection and disease processes needs to be used to estimate $\bs{\theta}$. 

The observed data $\boldsymbol{y}_t=(y_{t_u,\textrm{C}},y_{t_v,\textrm{G}}, y_{t_v,\textrm{P}},y_{t_k,\textrm{S}})$ include: $y_{t_u,\textrm{C}}$, the counts of confirmed cases during the initial weeks of the outbreak (Figure \ref{fig:obsFluTrans}(D)); $y_{t_v,\textrm{G}}$, the number of primary care consultations for ILI, including the individuals attending for non-pandemic ILI (Figure \ref{fig:obsFluTrans}(A)); $y_{t_v,\textrm{P}}$, the complementary virological data on nasopharyngeal positivity for A/H1N1 (Figure \ref{fig:obsFluTrans}(C) and Section \ref{sec:sevIntro}); and $y_{t_k,\textrm{S}}$, the cross-sectional sero-prevalence data (Figure \ref{fig:obsFluTrans}(B)). Typically there is some reporting delay between the disease diagnoses and their appearance in health-care surveillance, but for simplicity, here no such delay is assumed, so that $y_{t_i,i}$ is observed at the same time $t_i$ as the disease endpoint $N_{t_i,i}$ for each $i$. Each item of data informs ${\bs{\theta}}$ through a probabilistic link.

More specifically, $y_{t_k,\textrm{S}}$ is a realisation of a Binomial distribution,
$$
y_{t_k,\textrm{S}} \sim \textrm{Binomial}\left(n_{t_k,\textrm{S}}, \psi_{t_k,\textrm{S}} \right),
$$
with sample size $n_{t_k,\textrm{S}}$ and probability $\psi_{t_k,\textrm{S}}=1 - {\frac{S_{t_k}}{N}}$. The sero-prevalence data $y_{t_k,\textrm{S}}$ therefore directly inform the number of susceptibles $S_{t_k}$ and parameters $\bs\theta_T$ as $S_{t_k} = S_{t_k}({\bs\theta_T})$.

The counts of confirmed cases $y_{t_u,\textrm{C}}$ and ILI consultations $y_{t_v,\textrm{G}}$ are taken as realisations of negative binomial distributions, with means given by functional parameters $\psi_{t_u,\textrm{C}}$ and $\psi_{t_v,\textrm{G}}$ respectively, and time-varying over-dispersion parameter $\eta_t$, i.e.
$$
y_{t_v,G} \sim \textrm{Negative Binomial}\left(\psi_{t_v,G}, \eta_{t_v}\right)
$$
where
\begin{align*}
\psi_{t_u,\textrm{C}} & = N_{t_u,\textrm{C}} \\
\psi_{t_v,\textrm{G}} & = N_{t_v,\textrm{B}} + N_{t_v,\textrm{G}}
\end{align*}
and are, therefore, functions defined by convolution equations of the type in (\ref{eqn:convolution}).

To disentangle the GP ILI consultations due to the pandemic strain, $N_{t_v,\textrm{G}}$, from all other ILI consultations, $N_{t_v,\textrm{B}}$, information is needed on the proportion of all GP ILI consultations that result from the pandemic strain. This information is provided by virological positivity data, where observed positive samples $y_{t_v,\textrm{P}}$ are considered realisations of a Binomial distribution,
$$
y_{t_v,\textrm{P}} \sim \textrm{Binomial}\left(n_{t_v,\textrm{P}}, \psi_{t_v,\textrm{P}} \right),
$$
with sample size (number of tests) $n_{t_v,\textrm{P}}$ and probability parameter $\psi_{t_v,\textrm{P}} = 1 - \frac{N_{t_v,\textrm{B}}}{N_{t_v,\textrm{B}} + N_{t_v,\textrm{G}}} $. The proportion positive, $\psi_{t_v,\textrm{P}}$, is expressed as a function of the disentangled counts $N_{t_v,\textrm{B}}$ and $N_{t_v,\textrm{G}}$. As in Section \ref{sec:sevWave3}, the proportion positive $\psi_{t_v,\textrm{P}}$ and number of ILI consultations $\psi_{t_v,\textrm{G}}$ are jointly regressed on age and time, on logit and log scales respectively. The background counts $N_{t_v,\textrm{B}}$ are therefore a function of the regression parameters $\bs{\beta}_B$ (Figure \ref{fig:fluTransDAG}).

In each generic calendar time interval $[t_{j-1},t_j]$, as indicated in Section \ref{sec:evsyn}, the likelihood of the data is the product 
\begin{equation}\label{eqn:trans_likelihood}
L(\bs{y}_{t_j} \mid \bs{\theta}) = \prod_{i \in \{\textrm{C,G,P,S}\}} L(y_{t_j,i} \mid \bs{\theta}) 
\end{equation}
of the contributions $L(y_{t_j,i} \mid \bs{\theta})$ of the four data streams, as these are considered to be independent conditional on their common parents. The posterior distribution is then obtained by combining the likelihood with the prior distribution $p(\bs{\theta})$.

\subsection{Results}
\begin{figure}
\centering
\includegraphics[width = 0.9\textwidth]{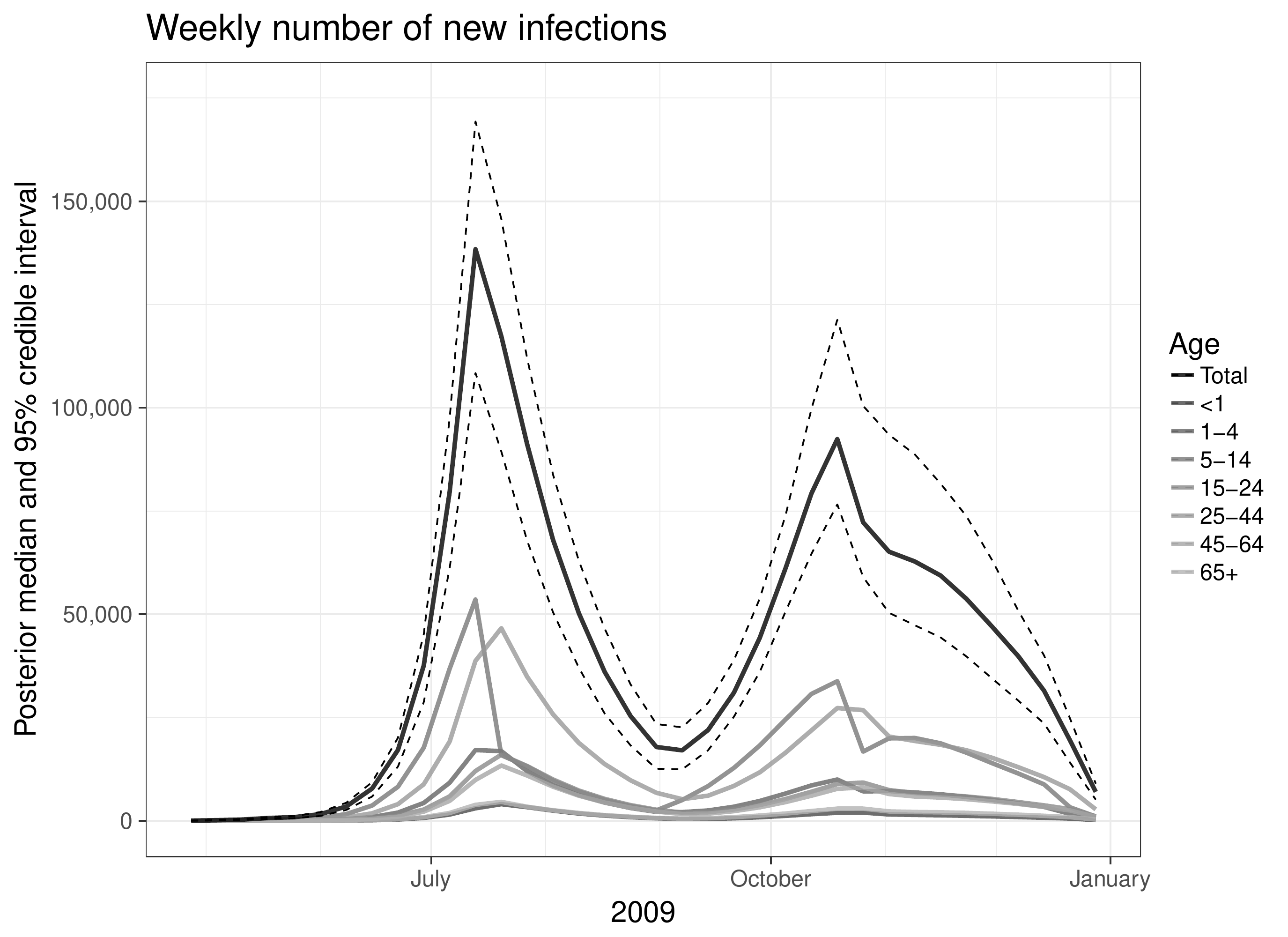}
\caption{Posterior median (solid lines) and 95\% credible interval (dashed) lines number of new infections per week in London, overall and by age group.}
\label{fig:fluSEIRpostr}
\end{figure}

Figure \ref{fig:fluSEIRpostr} shows the posterior distribution for the number of new A/H1N1 infections by age group in London, revealing the two waves of infection in summer and autumn/winter 2009. The first wave of infection has a higher peak, whereas the second wave, particularly by age group, is spread over a longer period of time, resulting in a higher attack rate in the second wave. $R_0$ is estimated to be 1.65 ($95\%$ credible interval 1.56-1.75).

\begin{figure}
\centering
\includegraphics[width = \textwidth]{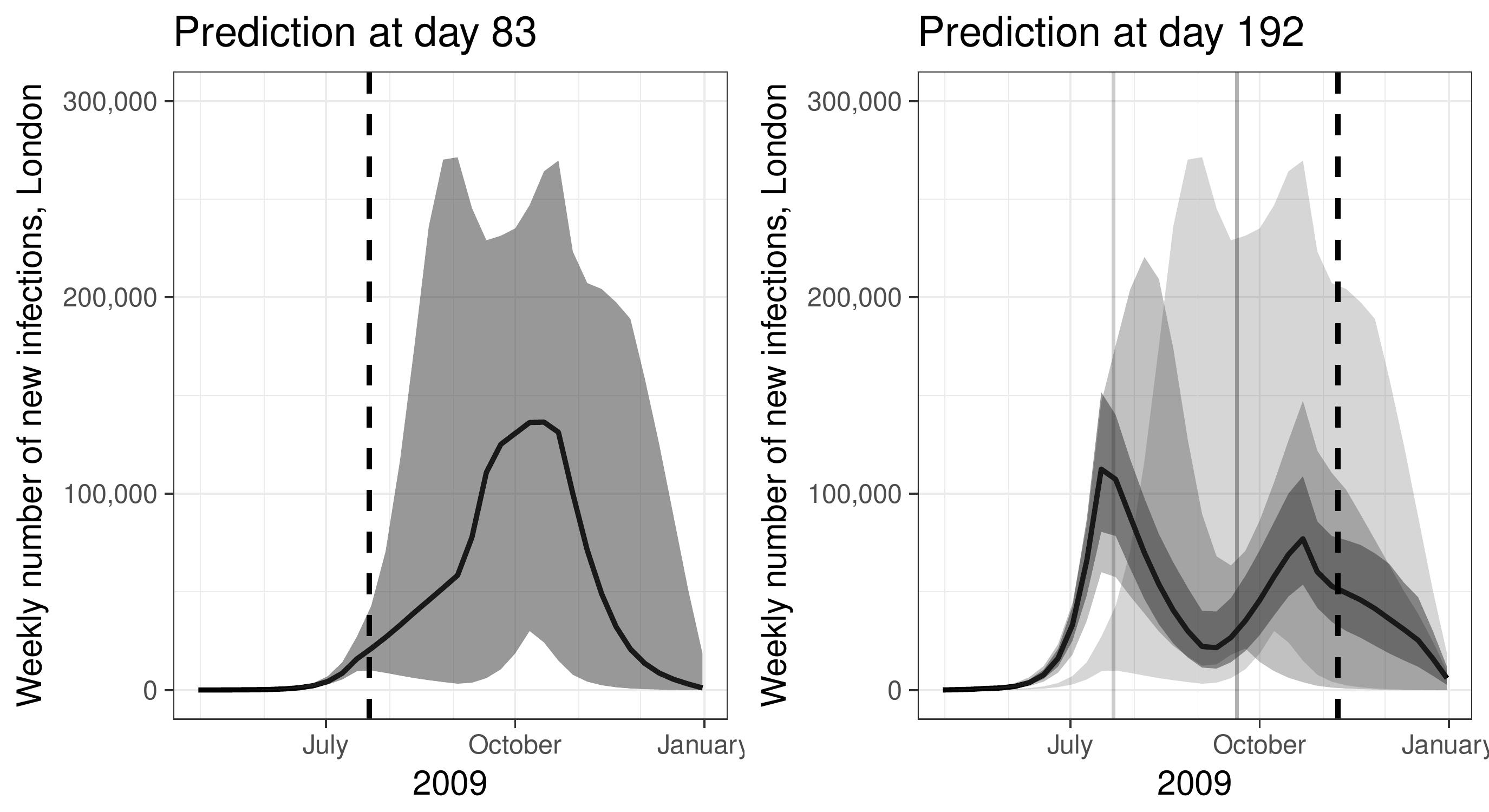}
\caption{Predictions of the weekly number of new infections in London, at days 83 and 192 of the epidemic (vertical dashed lines): posterior median (black line) and 95\% predictive intervals (shaded areas). For the day 192 prediction, paler shaded areas represent the predictive intervals at two previous timepoints (days 83 and 143 respectively, shown by the grey vertical lines). \label{fig:fluSEIRpredict}}
\end{figure}

Figure \ref{fig:fluSEIRpredict} shows predictions forward in time based on the data up to days 83 and 192 of the epidemic respectively. Note how the uncertainty in the predictions is progressively reduced as data accumulate.

\section{Model building, inference and criticism \label{sec:challenges}}

The two influenza case studies above, although relatively simple, have demonstrated how the combination of multiple sources can easily lead to complex probability models. This complexity can challenge standard inferential tools, motivating the development of novel approaches. In what follows, we continue to use the two influenza examples to illustrate such new approaches to model building, efficient inference and model criticism.

\subsection{Strategies for model building}

There are two strategies to build a complex evidence synthesis model: (i) all the data are combined simultaneously, as in joint models (e.g. \cite{Rizopoulos2012}); or (ii) the model is assembled in stages, using subsets of the evidence initially, before combining the results in a second stage, as in standard meta-analysis \cite{Borenstein2009}. When a model is complex, the latter strategy is sensible, to understand what might be inferred from each dataset in isolation. A staged approach is used in both the influenza examples of Sections \ref{sec:flusev} and \ref{sec:flutrans}. For example, in the severity model, a sub-model of a stochastic process describing entries to/exits from ICU is first fitted to the ICU prevalent case data in the third wave, before combining the results in the full evidence synthesis model. In the transmission model, a joint regression model of GP consultation and virological positivity data was fitted initially, before the results were incorporated in the transmission model to disentangle ``background'' non-influenza noise from the signal of influenza consultations.

\begin{figure}[h]
\centering
\includegraphics[width = 0.6\textwidth, trim = 500 600 550 100, clip]{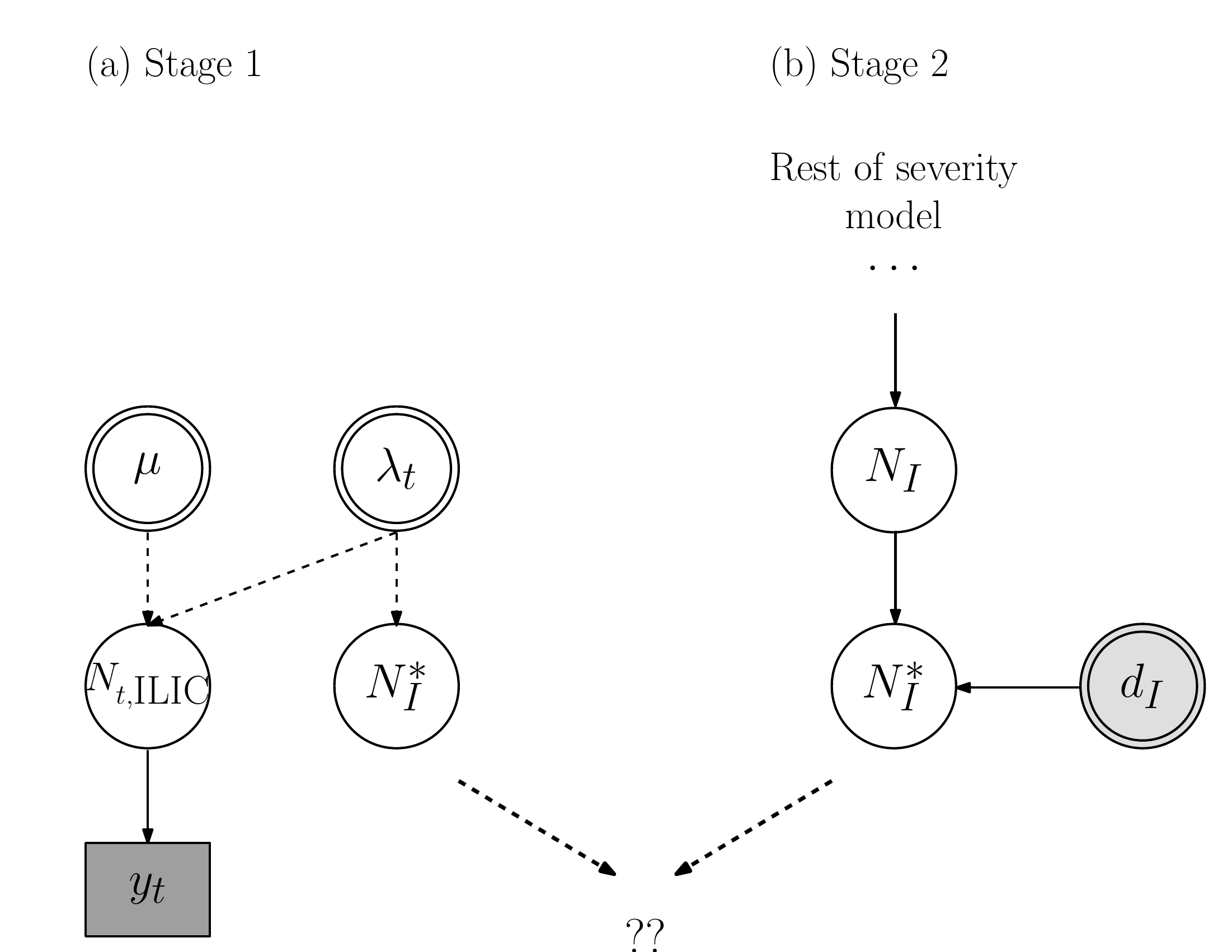}
\caption{Two-stage modelling strategy for joining ICU sub-model with
  rest of severity model.}
\label{fig:ICU2stage}
\end{figure}

To illustrate a two-stage process, we use the severity example. Figure \ref{fig:ICU2stage} shows a simplified schematic DAG of the stages: in both the stage 1 and stage 2 models, the cumulative number of ICU admissions measured over the time period of the ICU prevalence data, $N_{\textrm{I}}^*$, has a prior model. The stage 1 prior is in terms of the parameters of the ICU entry/exit process, $\lambda_t$ and $\mu$; whereas the stage 2 prior is in terms of the parameters of the period-prevalence-type severity model, i.e. the conditional and detection probabilities described in Section \ref{sec:flusev}. 

The existence of two prior models poses the question of how to combine the two sources of information. In \cite{Presanis2014}, the problem was solved using an approximate method, transferring the posterior mean (sd) estimate of $N_I^*$ from the stage 1 ICU sub-model to the stage 2 severity model via a likelihood term (equation (\ref{eq:ICUlogNorm}) and Figure \ref{fig:dagWave3}), such that the posterior of $\log(N_I^*)$ from the ICU sub-model is approximated by a Normal distribution. This approximate approach is acceptable when the approximation is good, i.e. when the sample size in the stage 1 model is large enough to guarantee a Gaussian posterior distribution. If not, in \cite{Goudie2016}, an alternative, more general, exact method for joining (and splitting) models, ``Markov melding'', is proposed.

In general terms, suppose we have $M$ probability submodels $p_m(\phi,\psi_m,Y_m)$, $m=1,\ldots,M$,  with parameters $\phi$ and $\psi_m$ and observable random variables $Y_m$. Suppose further that $\phi$ is common to all modules, acting as a `link' between the submodels, and that the aim is to combine all modules into a single model $p_{\mathrm{comb}}(\phi,\psi_1,\ldots,\psi_M,Y_1,\ldots,Y_M)$, so that the posterior distribution for the link parameter $(\phi)$ and the submodel-specific parameters $\psi_{m}$ reflects all information and uncertainty. In some cases, such model joining is readily achievable using standard hierarchical modelling constructs. However, there are some contexts where this joining is not straightforward, in particular where: (i) some sub-models may not be expressible conditional on the link parameter $\phi$, particularly if $\phi$ is a non-invertible deterministic function of other parameters in a sub-model; (ii) the prior marginal distributions $p_m(\phi)$ for $\phi$ differ in different submodels. Both of these situations arise in the severity estimation of Section \ref{sec:flusev} and Figure \ref{fig:ICU2stage}, where the link parameter $\phi$ is the cumulative number of ICU admissions $N_I^*$.

Markov melding \cite{Goudie2016} addresses model joining in these contexts by building on Markov combination \cite{DawidLauritzen1993,MassaLauritzen2010,MassaRiccomagno2017} and Bayesian melding \cite{PooleRaftery2000} ideas. Markov combination allows joining of sub-models under the restrictive constraint that the prior marginals $p_m(\phi)$ are identical for each $m$, {\it i.e.} the submodels $p_m(\phi,\psi_m,Y_m)$, $m=1,\ldots,M$ are {\it consistent} in the link parameter $\phi$: $p_m(\phi)= p(\phi)$  for all $m$. In practice, however, the marginal distributions $p_m(\phi)$ are usually \emph{not} exactly identical, as in Figure \ref{fig:ICU2stage}. Markov melding therefore exploits the Bayesian melding approach \cite{PooleRaftery2000} to replace each marginal with a pooled marginal distribution $$p_{pool}(\phi)=g(p_1(\phi),\ldots,p_M(\phi))$$, where $g$ is a pooling function chosen such that $\int g(\phi)\,d\phi = 1$ and $p_{pool}(\phi)$ is an appropriate summary of the individual marginals. Since each replaced model is now consistent in $\phi$, the Markov melded model can be obtained via a Markov combination of the replaced models:
\begin{equation*}
\begin{split}
p_{\textrm{MM}}(\phi,\psi_1,\ldots,\psi_M,Y_1,\ldots,Y_M)  
&=
p_{pool}(\phi)\prod_{m=1}^M\frac{ p_m(\phi,\psi_m,Y_m)}{p_m(\phi)}.
\end{split}
\end{equation*}

Different possible pooling functions $g$ for the melded marginal $p_{pool}(\phi)$ are discussed in \cite{Goudie2016}. Once the new model $p_{\textrm{MM}}$ has been formed, posterior inference given all the data $y_1,\ldots,y_M$ can be performed. Markov melding incorporates more data than any single submodel, and so will provide more precise inferences if the various components of evidence (priors and data) in each submodel do not substantially conflict. Otherwise, Markov melding may be misleading, so, before proceeding, the underlying reasons for the conflict should be investigated and resolved (see section \ref{sec:conflict}).

Note that the Markov melding method can, of course, be generalised to the case of multivariate link and submodel-specific parameters, $\bs{\phi}$ and $\bs{\psi}_m$ \cite{Goudie2016}.

\begin{figure}
    \centering
    \makebox{
    \includegraphics[width = \textwidth]{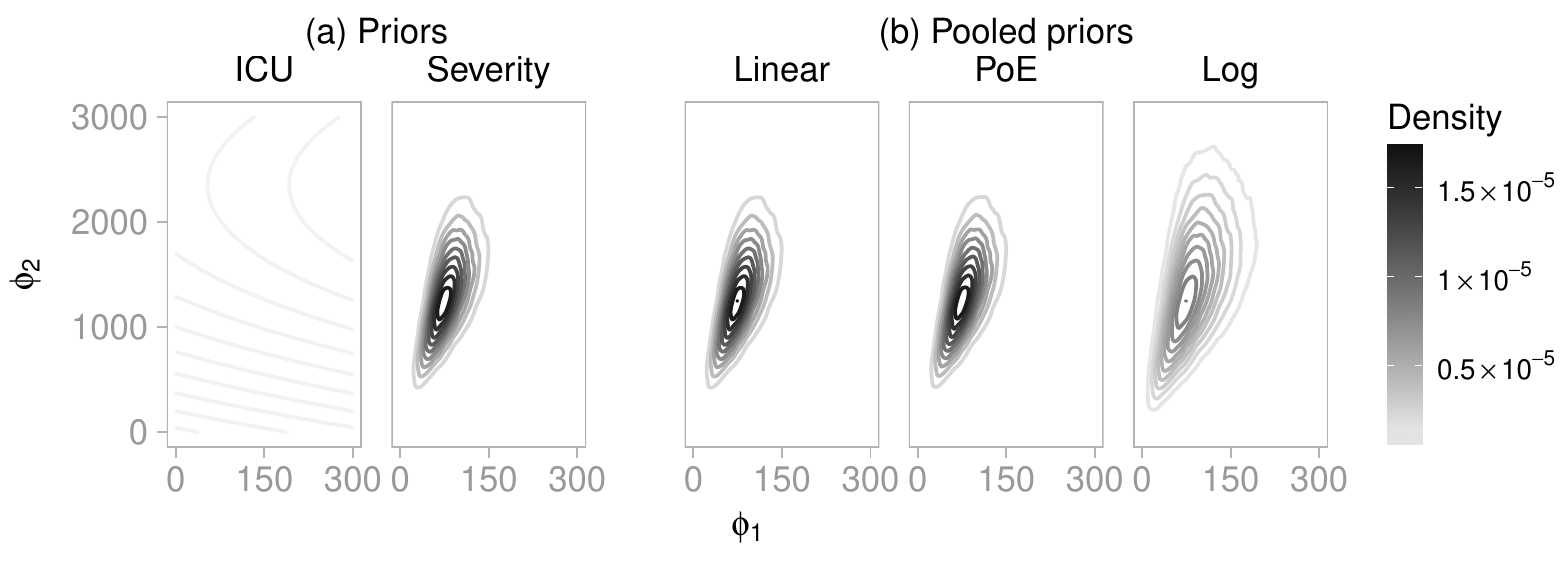}
    }
\caption{Pooled prior marginal distributions under different pooling
  functions for the ICU and severity sub-models. \label{fig:sevPoolPrior}}
  \end{figure}

\begin{figure}
    \centering
    \makebox{
    \includegraphics[width =0.7\textwidth]{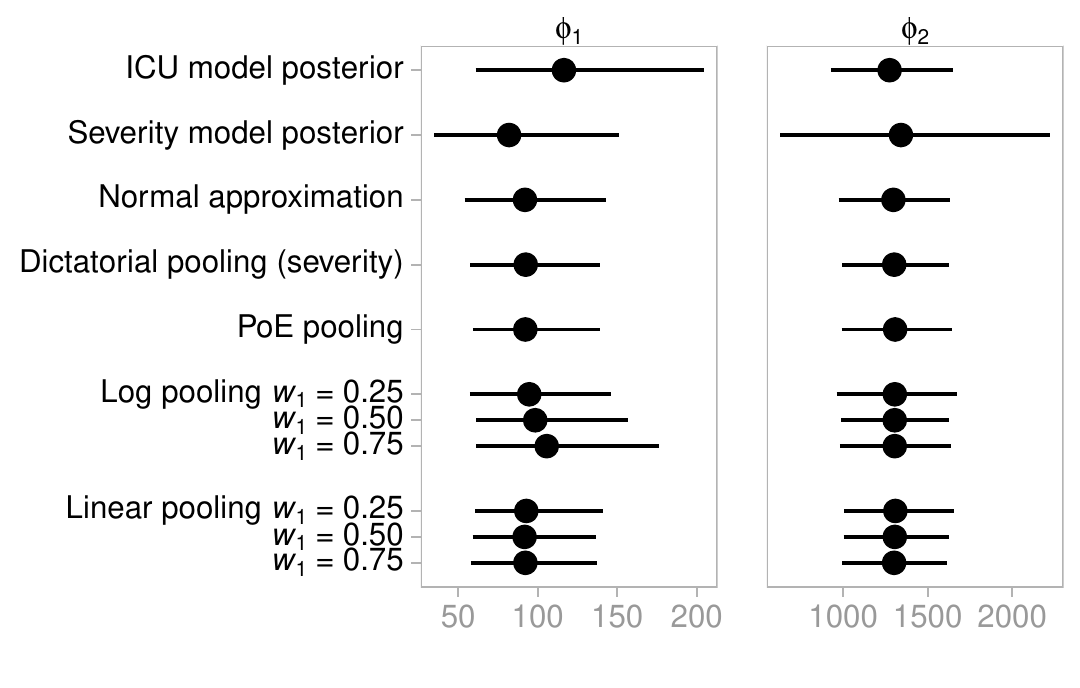}
    }
\caption{Posterior distributions (medians and $95\%$ credible intervals) for the ICU and severity model
  parameters under the Markov melded model with different pooling
  functions, in comparison to the separate
  sub-model posterior distributions. \label{fig:sevMeldPostr}}
  \end{figure}

\paragraph*{Markov melding in the influenza severity example}
For the severity example, the use of Markov melding on the ICU and severity sub-model implied priors (Figure \ref{fig:sevPoolPrior}) results in greater posterior precision under different pooling functions, in comparison to both the sub-model posteriors alone and the normal approximation employed in \cite{Presanis2014} (Figure \ref{fig:sevMeldPostr}).

\subsection{Computationally efficient inference}
Markov chain Monte Carlo (MCMC) methods \cite{GamermanLopes2006} have become standard tools in Bayesian inference to sample from posterior distributions. This sampling, in cross-sectional estimation problems of the type in Section \ref{sec:flusev} (e.g. \cite{McDonald2014}), can feasibly be carried out using available software implementing MCMC (e.g. \cite{Lunn2009}). On the other hand, inference for more complex models, such as transmission models similar to that of Section \ref{sec:flutrans} ({\it e.g.} \cite{Dorigatti2013}), requires bespoke code and a tailored MCMC algorithm. However, classical MCMC is often not a computationally viable option in stochastic transmission models when the data structure is complex ({\it e.g.} \cite{Shubin2016}) and can become computationally inefficient even in inference for deterministic models, when inferences are required within a restricted time frame. Alternative approaches to Bayesian inference are becoming popular,  often combined with MCMC, to tackle such computational challenges.  Examples include Approximate Bayesian Computation (ABC) ({\it e.g.} \cite{Ratmann2012}), used to estimate the likelihood; Sequential Monte Carlo (SMC) ({\it e.g.} \cite{Sheinson2014}), where inference is sequentially updated using only the most recent data; and emulation \cite{Farah2014} and history matching \cite{Andrianakis2017}, where a complex transmission model is replaced by a simpler approximate model.

As a concrete illustration, we re-visit the deterministic transmission model of Section \ref{sec:flutrans}, where the goal is to reconstruct retrospectively the evolution of the influenza A/H1N1 epidemic. In \cite{Birrell2011}, an adaptive Metropolis Hasting algorithm is used to derive the posterior distribution of $\bs\theta$ using 245 days of GP consultation data, combined with the various additional sources of information. To reconstruct the epidemic in London, each run of the MCMC algorithm requires around $7 \times 10^5$ iterations to reach convergence, taking more than four hours for a single MCMC chain The bottleneck is the evaluation of the likelihood in equation (\ref{eqn:trans_likelihood}), which involves calculation of the computationally expensive convolutions in equation (\ref{eqn:convolution}). This is an acceptable computational burden in the case of retrospective inference. However, in the midst of an epidemic, prospective estimation and prediction will be needed as new data arrive, for instance, daily. Then the MCMC algorithm would have to be rerun each day to re-analyse the complete dataset, which would not be optimal, particularly if alternative models need to be explored. More efficiently, the posterior distribution at each new time point could be derived from the one at the previous time point. In \cite{Birrell2016}, a hybrid SMC algorithm is developed to enable this more efficient use of the information and carry out inference and predictions in real time. The idea underlying SMC is to derive sequential posterior distributions $\pi_{k}(\bs\theta)=p(\bs\theta \mid {\bs y}_{{t_1}:{t_k}}) \propto p(\bs\theta)L({\bs y}_{{t_1}:{t_k}} \mid \bs\theta)$ for $k=1,\ldots, K$ as data accumulate. At each time point $t_k$ the distribution $\pi_{k}(\bs\theta)$ is approximated by $n_k$ particles $\{\bs\theta_k^{(1)},\ldots,\bs\theta_k^{(n_k)}\}$ with corresponding weights $\{\omega_{k}^{(1)},\ldots,\omega_{k}^{(n_k)}\}$. As data ${\bs y}_{t_{k+1}}$ arrive at $t_{k+1}$, $\pi_{k}(\bs\theta)$ serves as an importance distribution and the updated $\pi_{k+1}(\bs\theta)$ is obtained by re-weighting the sample $\{\bs\theta_k^{(1)},\ldots,\bs\theta_k^{(n_k)}\}$ by the importance ratios $\frac{\pi_{k + 1}\left(\bs\theta_k^{(j)}\right)}{\pi_k\left(\bs\theta_k^{(j)}\right)}$ for each $j \in 1, \ldots, n_k$. 

In the model of \cite{Birrell2016} this ratio reduces to the likelihood of the new data, so that the $j^{th}$ particle has weight 
\begin{equation*}
\omega_{k + 1}^{j} \propto \omega_{k}^{j} \frac{\pi_{k + 1}\left(\bs\theta_k^{(j)}\right)}{\pi_k\left(\bs\theta_k^{(j)}\right)} = \omega_{k}^{j}L(\bs y_{k+1} \mid {\bs \theta_{k}^{(j)}}).
\end{equation*}

This simple SMC scheme works well when the data follow a stable pattern, as demonstrated in settings where only one data stream is available (e.g. \cite{Ong2010}). However, in the specific application of \cite{Birrell2016}, the challenge is not particularly posed by the multiplicity of data, rather by the sudden change in the pattern of health seeking behaviour produced by a public health intervention (see Figure \ref{fig:obsFluTrans}(A)). Such a change introduces a shock to the system and complicates dramatically the tracking of the sequential distributions $\pi_k(\cdot)$ over time. On arrival of a particularly informative new batch of data ${\bs y}_{t_{k+1}}$, the sample $\{\bs\theta_k^{(1)},\ldots,\bs\theta_k^{(n_k)}\}$ degenerates to the few particles consistent with the new information, which, carrying large weights, give a misleading estimate of $\pi_{k+1}(\cdot)$. The naive SMC algorithm is then adapted to handle these highly informative observations by: introducing resampling and MCMC jittering steps \cite{GilksBerzuini2001} to rejuvenate the sample; and by sequentially including only fractions of the new data to minimise the divergence between posterior distributions at consecutive times \cite{DelMoral2006,Neal1996}. The result is a hybrid semi-automatic SMC algorithm that is more computationally efficient than the original MCMC, is highly parallelisable, and can deal with sudden shocks in the observational patterns.

\subsection{Model criticism: conflict and influence \label{sec:conflict}}

Model criticism is crucial to any analysis. However, specific to the context of multiple source evidence synthesis are: the potential for conflicting evidence, with such conflicts needing to be detected, quantified and resolved; and the critical assessment of what the role and influence of different sources is. 

In the influenza severity example of section \ref{sec:flusev}, the initial sensitivity analysis shown in the right-hand side of Figure \ref{fig:NsymByWaveAge} did not include a detection probability $d_{\textrm{S}}$ for the HPA estimates $\hat{y}_{\textrm{S}}$ of the number symptomatic (i.e. with $d_{\textrm{S}} = 1$). However, this ``naive'' model led to high posterior mean deviances, as shown in Table \ref{tab:devsNaiveSev} for `data' $\hat{y}_{\textrm{S}}$ on a log scale for
the first wave. 
\begin{table}
\centering
\caption{Deviance summaries for `data' on number symptomatic in first wave, log-scale, by age group, for the ``naive'' model: HPA estimate ($\hat{y}_{\textrm{S}}$) and corresponding standard deviation ($\hat{\sigma}_{\textrm{S}}$); posterior mean estimate ($N_{\textrm{S}}$) and corresponding $95\%$ credible interval (CrI); posterior mean deviance contributions ($\overline{D}$); plug-in deviance at posterior mean of parameters ($D(\overline{\theta})$); effective number
of parameters ($p_D$); deviance information criterion (DIC). \label{tab:devsNaiveSev}}
\begin{tabular}{rrrrrrrrrr}
\hline
Age & $\hat{y}_{\textrm{S}}$ & $\hat{\sigma}_{\textrm{S}}$ &$N_{\textrm{S}}$ & \multicolumn{2}{c}{$95\%$ CrI} & $\overline{D}$ & $D(\overline{\theta})$ & $p_D$ & DIC\\
\hline
$<1$  &  8.11 & 0.30 &  8.27 &  7.70 &  8.85 &   1.32 &   0.32 &  1.01 &   2.33 \\
1-4   &  7.81 & 0.26 &  9.89 &  9.42 & 10.38 &  66.72 &  65.80 &  0.92 &  67.65 \\
5-14  &  9.78 & 0.28 & 11.75 & 11.32 & 12.16 &  49.16 &  48.58 &  0.58 &  49.74 \\
15-24 & 10.23 & 0.26 & 11.17 & 10.74 & 11.62 &  14.05 &  13.30 &  0.76 &  14.81 \\
25-44 & 11.46 & 0.29 & 11.30 & 10.83 & 11.80 &   1.03 &   0.30 &  0.73 &   1.75 \\
45-64 & 12.03 & 0.26 & 10.06 &  9.61 & 10.52 &  59.17 &  58.36 &  0.81 &  59.98 \\
65+   & 11.25 & 0.27 &  7.62 &  7.12 &  8.15 & 175.41 & 174.49 &  0.92 & 176.33 \\
\hline
\end{tabular}
\end{table}
By comparison, the model assuming the HPA estimates are under-estimates has much lower posterior mean deviances and DIC contributions (Table \ref{tab:devsSev}). The lack of fit in the ``naive'' model motivated a closer look at the consistency of the different sources of evidence about the denominators, or infections at lower levels of severity (asymptomatic and symptomatic), resulting in both the sensitivity analyses of \cite{Presanis2014} and more formal conflict assessment in \cite{Presanis2013}.
\begin{table}
\centering
\caption{Deviance summaries for `data' on number symptomatic in first wave, log-scale, by age group, for the model assuming HPA estimates are under-estimates. \label{tab:devsSev}}
\begin{tabular}{rrrrrrrrrr}
\hline
Age & $\hat{y}_{\textrm{S}}$ & $\hat{\sigma}_{\textrm{S}}$ &$N_{\textrm{S}}$ & \multicolumn{2}{c}{$95\%$ CrI} & $\overline{D}$ & $D(\overline{\theta})$ & $p_D$ & DIC\\
\hline
$<1$  &  8.11 & 0.30 &  8.11 &  7.56 &  8.66 & 0.89 & 0.00 & 0.89 & 1.78 \\
1-4   &  7.81 & 0.26 &  9.71 &  9.20 & 10.23 & 0.92 & 0.06 & 0.85 & 1.77 \\
5-14  &  9.78 & 0.28 & 11.45 & 10.93 & 11.95 & 0.82 & 0.00 & 0.82 & 1.64 \\
15-24 & 10.23 & 0.26 & 11.11 & 10.64 & 11.58 & 1.00 & 0.23 & 0.77 & 1.76 \\
25-44 & 11.46 & 0.29 & 11.22 & 10.71 & 11.72 & 0.90 & 0.00 & 0.90 & 1.80 \\
45-64 & 12.03 & 0.26 &  9.91 &  9.41 & 10.41 & 0.86 & 0.00 & 0.86 & 1.73 \\
65+   & 11.25 & 0.27 &  7.56 &  7.03 &  8.10 & 0.92 & 0.00 & 0.92 & 1.84 \\
\hline
\end{tabular}
\end{table}

\paragraph*{Conflict assessment methods}
Bayesian predictive diagnostics (e.g. \cite{Box1980,Rubin1984,Gelman1996,BayarriCastellanos2007}) have long been used in model assessment, comparing observations to predictions from the model. Posterior predictive tests \cite{GelmanBook2003} are known to be conservative, due to using the data both to fit the model and to compare to model predictions, so a variety of (computationally intensive) post-processing, approximate, or cross-validatory methods have been proposed instead (e.g. \cite{BayarriBerger2000,BayarriCastellanos2007,Hjort2006,SteinbakkStorvik2009,Gelman1996,MarshallSpiegelhalter2007}). Typically, each of these methods have been employed to assess models of a single dataset, rather than for an evidence synthesis.

``Conflict p-values'' \cite{MarshallSpiegelhalter2007,GasemyrNatvig2009,Presanis2013,Gasemyr2016} have been proposed as a generalisation of Bayesian cross-validatory predictive p-values that compare not only subsets of data to predictions resulting from the rest of the data, but also whole sub-models, comprising data, model structure and prior information, with predictions from the rest of the model. The key idea,  known as ``node-splitting'', is to split a DAG $\mathcal{G}(\phi, \bs{\theta}_{\setminus \phi}, \bs{y})$, comprising data $\bs{y}$ and latent quantites $\bs{\theta} = (\phi, \bs{\theta}_{\setminus \phi})$, into two independent partitions at any ``separator'' node $\phi$, $\mathcal{G}(\phi_a, \bs{\theta}_{a\setminus \phi}, \bs{y}_a)$ and $\mathcal{G}(\phi_b, \bs{\theta}_{b\setminus \phi}, \bs{y}_b)$. Two copies of the separator $\phi$ are created, $\phi_a$ and $\phi_b$, that are each identifiable in partitions $\mathcal{G}(\phi_a, \bs{\theta}_{a\setminus \phi}, \bs{y}_a)$ and $\mathcal{G}(\phi_b, \bs{\theta}_{b\setminus \phi}, \bs{y}_b)$ respectively. The aim is to compare the posterior distributions from each partition, $p(\phi_a \mid \bs{y}_a)$ and $p(\phi_b \mid \bs{y}_b)$.

This comparison is achieved by defining a difference function $\delta = h(\phi_a) - h(\phi_b)$, on an appropriate scale $h(\cdot)$, and considering where $0$ lies in the posterior distribution of the difference, $p_{\delta}(\delta \mid \bs{y}_a, \bs{y}_b)$. A two-sided conflict p-value corresponding to the hypothesis test $H_0: \delta = 0$ is defined as $c = 2 \times \min\left[ \textrm{Pr} \left\{ p_{\delta}(\delta \mid \bs{y}_a, \bs{y}_b) < p_{\delta}(0 \mid \bs{y}_a, \bs{y}_b) \right\}, 1 - \textrm{Pr} \left\{ p_{\delta}(\delta \mid \bs{y}_a, \bs{y}_b) < p_{\delta}(0 \mid \bs{y}_a, \bs{y}_b) \right\} \right]$, with different methods for evaluating $c$ and one-sided variations given in \cite{GasemyrNatvig2009,Presanis2013}. The conflict p-value has been demonstrated to be uniform under the null model in a range of models by \cite{Gasemyr2016}. This setup can be generalised to multiple partitions and multiple node-splits \cite{Presanis2017}.

\begin{figure}[h]
\centering
\includegraphics[width = \textwidth, trim = 30 600 300 100, clip]{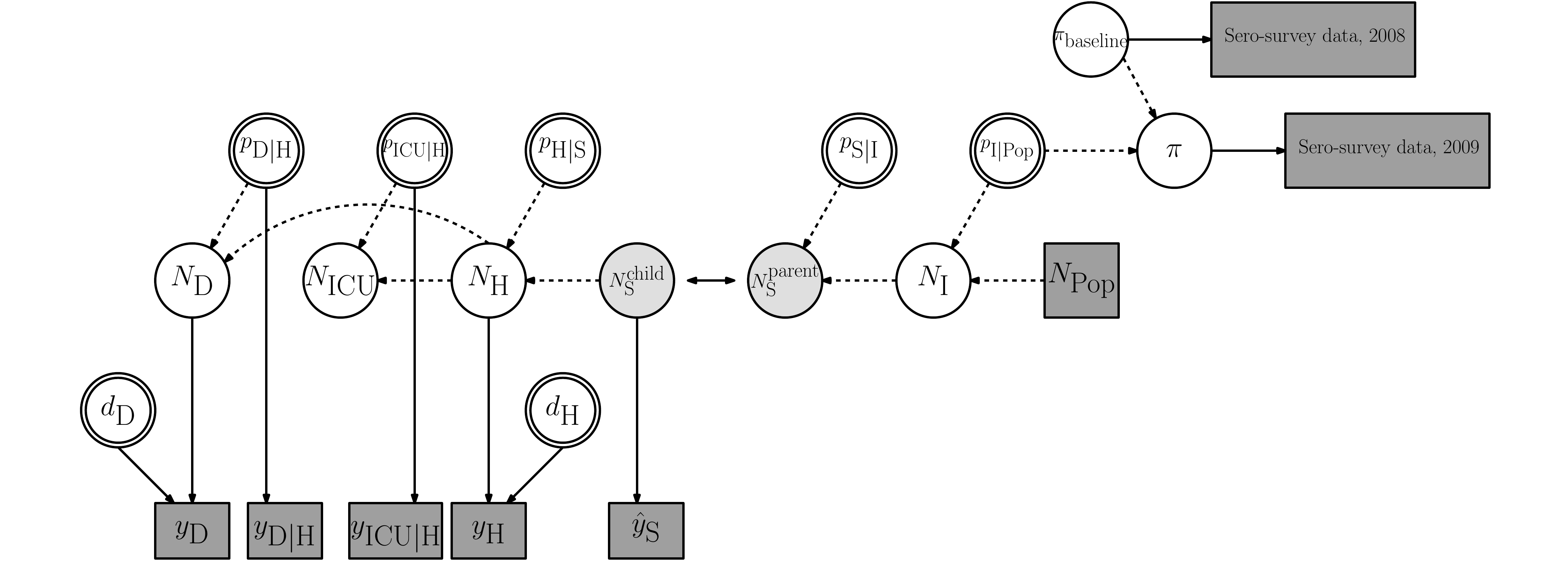}
\caption{DAG showing node split at $N_{\textrm{S}}$. On the right is the ``parent'' model and on the left the ``child'' model. The double-headed arrow represents the comparison between the two.}
\label{fig:fluDAGsplitNs}
\end{figure}

\begin{figure}[h]
\centering
\includegraphics[width = 0.9\textwidth]{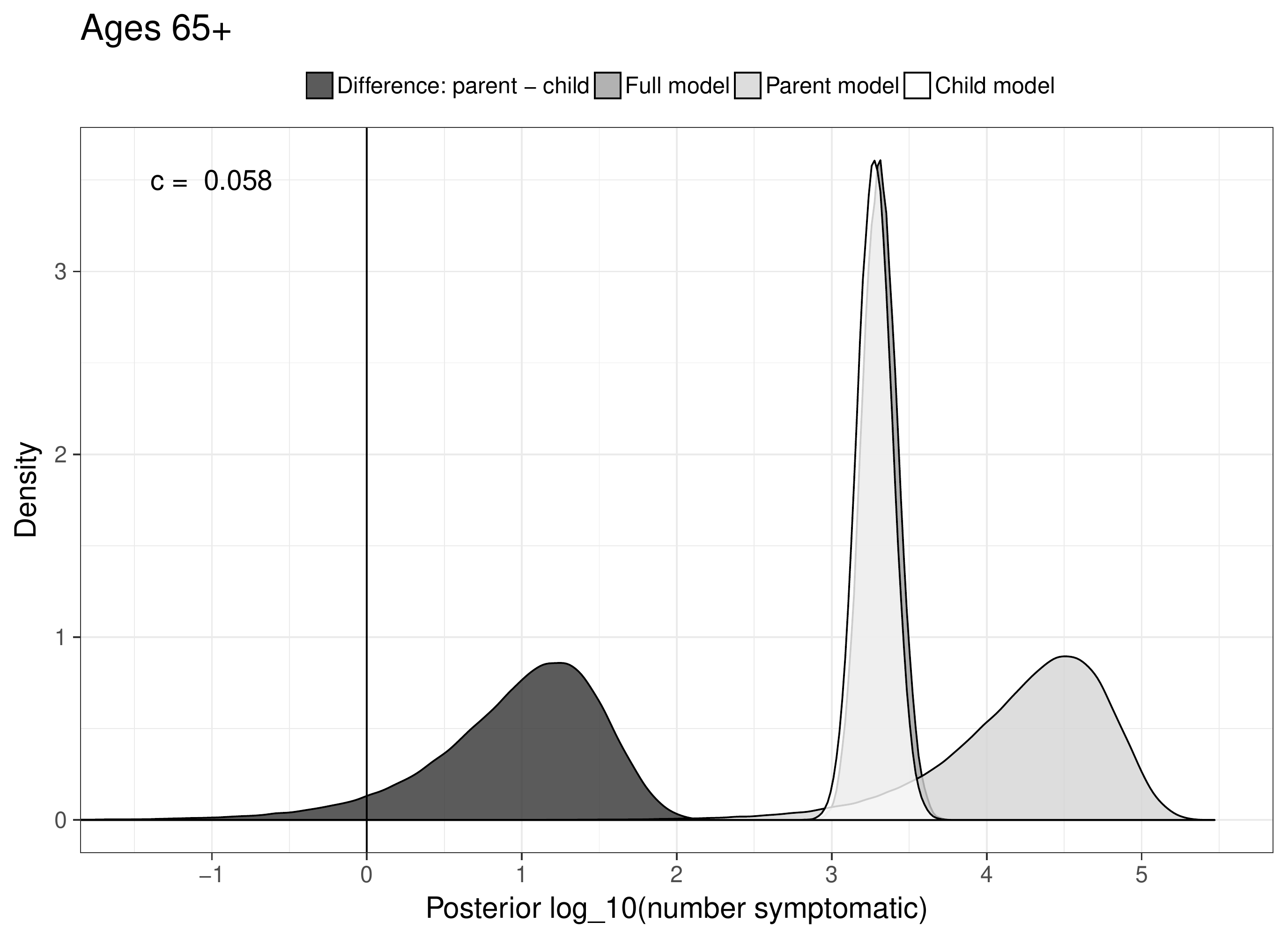}
\caption{Conflict (posterior difference) at $N_{\textrm{S}}$ between the parent and child models in age group 65+ in the first wave: difference function $\delta = \log_{10}(N_{\textrm{S}}^{\textrm{parent}}) - \log_{10}(N_{\textrm{S}}^{\textrm{child}})$ (dark grey); $\log_{10}(N_{\textrm{S}})$ from full model (medium grey); $\log_{10}(N_{\textrm{S}}^{\textrm{parent}})$ from parent model (light grey); $\log_{10}(N_{\textrm{S}}^{\textrm{child}})$ from child model (ivory).}
\label{fig:fluDelta65}
\end{figure}

\paragraph*{Conflict in the influenza severity example}
To assess whether the lack of fit to the HPA estimates of the number symptomatic in the ``naive'' model (Table \ref{tab:devsNaiveSev}) could be due to conflicting evidence, the model for the first wave only is split into two partitions at $N_{\textrm{S}}$, as in Figure \ref{fig:fluDAGsplitNs}. The ``parent'' partition comprises the data and priors informing the parent nodes of $N_{\textrm{S}}$, i.e. the sero-prevalence data and an informative prior for the proportion of infections that are symptomatic, $p_{\textrm{S} \mid \textrm{Inf}}$. The rest of the evidence (the severe case data and priors) comprises the ``child'' model.
The test for conflict between the parent and child models demonstrated low posterior probabilities of no conflict, particularly for the youngest and oldest age groups \cite{Presanis2013}. The posterior difference function $\delta = \log_{10}(N_{\textrm{S}}^{\textrm{parent}}) - \log_{10}(N_{\textrm{S}}^{\textrm{child}})$ is plotted for the age group $65+$ in Figure \ref{fig:fluDelta65}, together with the conflict p-value $c = 0.058$ and the corresponding posterior distributions for the two partitions and the full model. Note that the sero-prevalence data in the parent model imply a much higher, though also much more uncertain, number symptomatic than the severe case data in the child model. The lack of certainty in the sero-prevalence data means the severe case data and priors have more influence in the full model, so the full model posterior is much closer to the child model posterior than the parent one.

\begin{figure}[h]
\centering
\includegraphics[width = 0.9\textwidth]{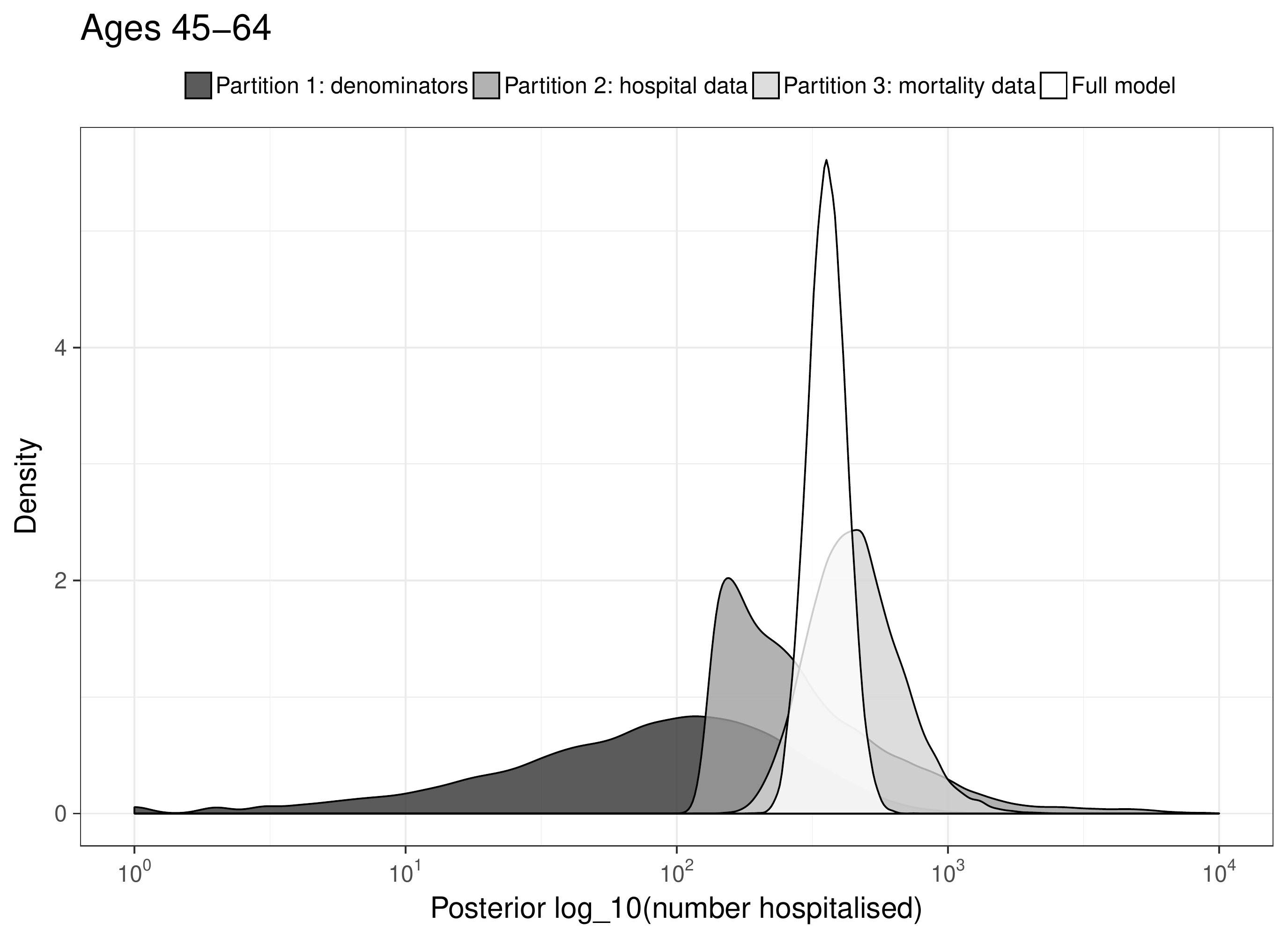}
\caption{Influence and conflict at $N_{\textrm{H}}$ between the three partitions in age group 45-64 in the first wave.}
\label{fig:fluInfluenceNh}
\end{figure}

A similar investigation of conflict at a different node in the ``naive'' model, the number of hospitalisations $N_{\textrm{H}}$, leads to splitting the DAG into three partitions (not shown), based on: the sero-prevalence data; the hospital data; and the mortality data, respectively. The influence of the evidence in the three partitions on the full model is shown in Figure \ref{fig:fluInfluenceNh}.
As in the previous example, the sero-prevalence (denominator) data in partition 1 are the most uncertain, with the hospital data partition (2) having less uncertainty, and the mortality data partition (3) even less uncertainty. The posterior distributions for all three partitions overlap substantially however, so that conflict is not detectable. Instead, when the three partitions are combined in the full model, we obtain a much more precise estimate of the number hospitalised, which is a compromise between the three partitions, as would be expected.

\section{Discussion \label{sec:discuss}}

This chapter has illustrated both the advantages and complexities of synthesising multiple data sources to estimate various hidden characteristics of infectious disease, through the two examples of severity and transmission estimation for influenza. Section \ref{sec:challenges} introduced three sets of tools to approach the challenges of complex model building, computationally efficient inference and model criticism respectively. However, these tools are a first step in resolving these challenges, with a number of questions remaining open.

Markov melding generalises existing ideas that facilitate realistic evidence synthesis, via a modular approach to model building. Some outstanding challenges include: the choice of pooling function; and the degree of heterogeneity between prior models that is acceptable for Markov melding to be appropriate -- when does heterogeneity become conflict? Nevertheless, Markov melding is an important step in this era of big data, generalising divide-and-conquer approaches \cite{Bardenet2017}.

Efficient inference for transmission models has been introduced using a sequential approach. Currently, the sequential approach is highly tailored to accommodate shocks to the system, so a question of interest is how to adapt a transmission model structure in real-time, to be able to generalise the SMC approach to any plausible epidemic scenario. The influenza model illustrated is deterministic in its dynamics, and the generic multi-state model description in Section \ref{sec:evsyn} and Figure \ref{fig:dagMSM}, while in theory accommodating stochastic dynamic transmission, is more focussed on deterministic dynamics. An important area of research is to consider efficient, real-time, inference from multiple sources for stochastic epidemics, particularly in the early stages of an outbreak \cite{Birrell2017}. The influenza transmission model of \cite{Shubin2016} includes several levels of stochasticity and data, but is highly computationally intensive, and therefore not feasible in real-time. How much stochasticity is therefore necessary to realistically model an emerging outbreak? 

The illustration of conflict assessment in Section \ref{sec:conflict} is targeted, in that conflict was assessed at particular nodes in the DAG of the influenza severity model, following suspected biases in particular data sources. However, in some contexts, it may not be so clear where to look for potential conflict, in which case a systematic search throughout a DAG for conflict may be warranted. However, such systematic assessment entails multiple tests, either through a multivariate difference function (as in the example of Figure \ref{fig:fluInfluenceNh}) or through fitting multiple node-split models. A framework for systematic assessment, accounting for the multiple tests and their correlation, has therefore been proposed \cite{Presanis2017}. Further open questions in this area include how to improve power to detect conflict, and how to make such methods more accessible by improving the computational feasibility of systematic conflict assessment. As with any cross-validatory framework, multiple node-splitting can be computationally burdensome, so for hierarchical models, \cite{Ferkingstad2017} have proposed an INLA approach to fast conflict diagnostics. A final area of open research related to understanding the influence of different, potentially conflicting, evidence sources on inference is the adaptation of value of information methods to evidence synthesis \cite{Jackson2017}.

\section*{Acknowledgements}

The authors would like to thank Robert Goudie and Paul Birrell for their input, and Johannes Bracher and Leonhard Held for constructive feedback. Daniela De Angelis and Anne Presanis are supported by the Medical Research Council (Unit programme number MC\_UU\_00002/11).

\bibliographystyle{apalike}
\bibliography{bibtex_ch5}
\end{document}